\documentclass[english,aps,preprintnumbers,aps,superscriptaddress]{revtex4-1}
\usepackage[T1]{fontenc}
\usepackage[latin9]{inputenc}
\usepackage{verbatim}
\usepackage{mathrsfs}
\usepackage{amsmath}
\usepackage{amssymb}
\usepackage{graphicx}
\usepackage[colorlinks=true,linktocpage=true,linkcolor=blue,citecolor=blue]{hyperref}
\usepackage{babel}
\usepackage{color}

\begin{document}

\title{From Kadanoff--Baym to Boltzmann equations for massive spin-1/2 fermions}

\author{Xin-Li Sheng}

\affiliation{Key Laboratory of Quark and Lepton Physics (MOE) and Institute of Particle Physics, Central China Normal University, Wuhan, 430079, China}

\author{Nora Weickgenannt}

\affiliation{Institute for Theoretical Physics, Goethe University, Max-von-Laue-Str.\ 1,
D-60438 Frankfurt am Main, Germany}

\author{Enrico Speranza}

\affiliation{Illinois Center for Advanced Studies of the Universe and Department of Physics, University of Illinois at Urbana-Champaign, Urbana, IL 61801, USA}

\affiliation{Helmholtz Research Academy Hesse for FAIR, Campus Riedberg, Max-von-Laue-Str.\ 12,
D-60438 Frankfurt am Main, Germany}

\author{Dirk H.\ Rischke}

\affiliation{Institute for Theoretical Physics, Goethe University, Max-von-Laue-Str.\ 1,
D-60438 Frankfurt am Main, Germany}

\affiliation{Helmholtz Research Academy Hesse for FAIR, Campus Riedberg, Max-von-Laue-Str.\ 12,
D-60438 Frankfurt am Main, Germany}

\affiliation{Interdisciplinary Center for Theoretical Study and Department of Modern Physics, University of Science and Technology of China, Hefei, Anhui 230026, China}

\author{Qun Wang}

\affiliation{Interdisciplinary Center for Theoretical Study and Department of Modern Physics, University of Science and Technology of China, Hefei, Anhui 230026, China}
\affiliation{Peng Huanwu Center for Fundamental Theory, Hefei, Anhui 230026, China}

\begin{abstract}
We derive Boltzmann equations for massive spin-1/2 fermions with local and nonlocal
collision terms from the Kadanoff--Baym equation in the Schwinger--Keldysh formalism,
properly accounting for the spin degrees of freedom. The Boltzmann equations are
expressed in terms of matrix-valued spin distribution functions, which are
the building blocks for the quasi-classical parts of the Wigner functions.
Nonlocal collision terms appear at next-to-leading order in $\hbar$ and
are sources for the polarization part of the matrix-valued spin distribution functions.
The Boltzmann equations for the matrix-valued spin
distribution functions pave the way for simulating spin-transport processes
involving spin-vorticity couplings from first principles.
\end{abstract}

\preprint{USTC-ICTS/PCFT-21-12}

\maketitle

\section{Introduction}

The goal of high-energy heavy-ion collisions is to create
a new state of strong-interaction matter, called the quark-gluon plasma (QGP),
at extreme conditions of temperature and density, and to study its properties
\citep{Rischke:2003mt,Gyulassy:2004zy,Shuryak:2004cy}.
In this state, the fundamental degrees of freedom of quantum chromodynamics (QCD),
quarks and gluons, are deconfined. The
collective flow observed in experiments \citep{Ackermann:2000tr}
reveals that the QGP is a nearly perfect fluid with a very low ratio of shear viscosity
to entropy density \citep{Kovtun:2004de}. The collective flow can
be quantitatively described by relativistic hydrodynamical models
\citep{Kolb:2003dz,Heinz:2013th,Florkowski:2017olj,Romatschke:2017ejr},
which are by now a well-established tool to describe the evolution of QCD matter
in heavy-ion collisions. However, until very recently these models did not
account for the dynamics of the spin degrees of freedom.

Noncentral heavy-ion collisions have a large orbital angular momentum (OAM),
which may polarize the spin of particles in strong-interaction matter in a way similar
to the well-known Barnett effect \citep{Barnett:1935}. A global polarization of particles
created in high-energy heavy-ion collisions was first proposed as a result of the spin-orbit
coupling \citep{Liang:2004ph,Liang:2004xn}
[see also Refs.\ \citep{Voloshin:2004ha,Betz:2007kg,Becattini:2007sr}].
In 2007, the STAR collaboration measured the global polarization of
$\Lambda$ hyperons in Au+Au collisions at 200 GeV, but the result
was zero within errors \citep{Abelev:2007zk}. A nonvanishing global
polarization of $\Lambda$ hyperons was measured by the STAR collaboration
in Au+Au collisions at lower energies \citep{STAR:2017ckg} and at
200 GeV with high precision \citep{Adam:2018ivw} [see, e.g., Refs.\
\citep{Wang:2017jpl,Becattini:2020ngo,Gao:2020vbh} for recent reviews].

The experimental data for the global polarization can be described
by theoretical models
\citep{Karpenko:2016jyx,Xie:2017upb,Li:2017slc,Sun:2017xhx,Wei:2018zfb}.
All these models feature a spin-vorticity coupling resulting from
microscopic spin-orbit coupling. Here the vorticity is the result
of the local rotation of the fluid \citep{Baznat:2013zx,Csernai:2013bqa,Csernai:2014ywa,Becattini:2015ska,Teryaev:2015gxa,Jiang:2016woz,Deng:2016gyh,Ivanov:2017dff,Shi:2017wpk}.
Statistical models for relativistic fluids consisting of particles with spin
\citep{Becattini:2007nd,Becattini:2007sr,Becattini:2013fla}
can be constructed based on the maximum-entropy principle
\citep{Zubarev_tmp1979_zps,Weert_ap1982,Becattini:2014yxa}.
A lot of progress has been recently made in theoretical studies of spin
polarization and vorticity formation in heavy-ion collisions. The
theoretical models can be grouped into two main categories: microscopic
and macroscopic models. Microscopic models are based on scatterings
of particles with spin-orbit coupling \citep{Gao:2007bc,Chen:2008wh,Huang:2011ru}.
Macroscopic models mainly include statistical models
\citep{Becattini:2007nd,Becattini:2007sr,Becattini:2013fla}
for fluids consisting of particles with spin in global equilibrium
\citep{Becattini:2009wh,Becattini:2012tc,Becattini:2015nva,Hayata:2015lga}
and spin-hydrodynamical models
\citep{Florkowski:2017ruc,Florkowski:2017dyn,Montenegro:2017lvf,Montenegro:2017rbu,Hattori:2019lfp,Gallegos:2021bzp,Li:2020eon,Bhadury:2020puc,Fukushima:2020ucl}
[see Ref.\ \citep{Florkowski:2018fap,Speranza:2020ilk} for recent reviews]. Kinetic
theory based on the Wigner-function formalism
\citep{Gao:2012ix,Chen:2012ca,Hidaka:2016yjf,Gao:2017gfq,Gao:2018wmr,Huang:2018wdl,Carignano:2018gqt,Liu:2018xip,Gao:2019zhk,Yang:2020mtz,Hou:2020mqp}
can also describe the spin polarization for massive fermions determined
by the axial-vector or tensor component of the Wigner function
\citep{Fang:2016vpj,Weickgenannt:2019dks,Gao:2019znl,Hattori:2019ahi,Wang:2019moi,Liu:2020flb}.
However, the above works do not include particle collisions, which are, however,
necessary in order to describe the spin dynamics of massive fermions.

In statistical models and spin-hydrodynamical models the spin-polarization
effect is described by the spin potential $\Omega_{\mu \nu}$ coupled to the
spin tensor $\Sigma^{\mu \nu}$ in the density matrix. However, unless one considers
a global-equilibrium state or a special choice for the energy-momentum tensor,
the spin potential is a priori not related to the thermal vorticity
$\varpi_{\mu \nu} \equiv - \frac{1}{2}
\left(\partial_\mu \beta_\nu - \partial_\nu \beta_\mu\right)$, where
$\beta^\mu \equiv u^\mu /T$, with $u^\mu$ the 4-velocity of matter and
$T$ the temperature \citep{Florkowski:2017ruc,Becattini:2018duy,Florkowski:2018ahw}.
Therefore, the main question is how the spin potential evolves to its
global-equilibrium value, given by a constant thermal vorticity.
A Boltzmann equation with a nonlocal collision term is essential
to describe such a dynamical process. A local collision term with spin degrees
of freedom has been studied under some approximations in
Refs.\ \citep{Li:2019qkf,Kapusta:2020npk}.
In Ref.\ \citep{Zhang:2019xya}, a microscopic model for spin polarization
through spin-orbit coupling in particle collisions was considered, which is
based on collisions of partons as wave packets. The method of wave
packets is an effective way of dealing with particle scatterings at
nonvanishing impact parameter. The spin-vorticity coupling naturally
emerges from the spin-orbit one encoded via polarized scattering amplitudes
in the collision integrals. Such a microscopic model provides a transparent
picture for the way how spin polarization can arise from vorticity.
There is only one missing piece in the model: the back reaction is
not considered, which converts spin into vorticity. Thus, this model does
not reach a state where spin is equilibrated. A first attempt to systematically incorporate
nonlocal collisions in a kinetic framework based on quantum field
theory was recently made in Ref.\ \citep{Yang:2020hri}, however, without
giving an explicit expression for the collision term at order $\mathcal{O}(\hbar)$. In
previous work \citep{Weickgenannt:2020aaf,Weickgenannt:2021cuo}, we derived the
collision term to order $\mathcal{O}(\hbar)$ in the Boltzmann equation
for massive spin-1/2 particles
in the Wigner-function approach \citep{DeGroot:1980dk}. The nonlocality of the
collision term allows for the conversion of
orbital into spin angular momentum. We showed that the collision term vanishes
in global equilibrium and that the spin potential is then equal to a constant value of the
thermal vorticity.

In this paper, we derive Boltzmann equations for massive spin-1/2 fermions
in the Schwinger-Keldysh, or closed-time-path (CTP), formalism
\citep{Martin:1959jp,Keldysh:1964ud} with collision terms of leading
and next-to-leading order in $\hbar$ [for reviews of the CTP formalism, see, e.g.,
Refs.\ \citep{Chou:1984es,Blaizot:2001nr,Berges:2004yj},
for a recent application of the CTP formalism to dissipative hydrodynamics, see
Ref.\ \citep{Crossley:2015evo}].
Our goal is to derive Boltzmann equations that are suitable for
the simulation of spin transport processes involving the spin-vorticity coupling.

We consider a system of massive spin-1/2 fermions interacting
via generic one-boson exchange. We will assume the interaction range to be much
smaller than all other scales in the problem, which effectively reduces
the interaction to a four-fermion vertex, similar to the time-honored
Nambu--Jona-Lasinio (NJL) model \citep{Nambu:1961tp,Nambu:1961fr}.

The paper is organized as follows. In Sec.\ \ref{sec:qed-in-ctp} the model system is
introduced and a short introduction to the CTP formalism is given.
In Sec.\ \ref{sec:kb} we derive the Kadanoff--Baym (KB)
equation \citep{Kadanoff:1962} in quasi-particle approximation
for the Wigner function from the Dyson-Schwinger
equation on the CTP contour. We decompose the
Wigner function in terms of its Clifford components and derive a
system of equations of motion for the latter. We also derive
mass-shell conditions and Boltzmann-type equations for the
latter and show that off-shell contributions cancel to lowest
order in the coupling constant. In Sec.\ \ref{sec:h-expand-kb} we perform an
expansion of the KB equation in powers of Planck's constant $\hbar$.
The structure of the matrix-valued spin distribution functions
and the quasi-classical parts of the Wigner functions are discussed in
Sec.\ \ref{sec:spin-dist}.
The Boltzmann equations for the matrix-valued spin distribution functions
at leading and next-to-leading order in $\hbar$ are derived in
Secs.\ \ref{sec:boltzmann-leading}
and \ref{sec:boltzmann-next}, respectively.
We close this work with a summary of the results in Sec.\ \ref{sec:summary}.

We adopt the following notation and conventions:
$a\cdot b=a^{\mu}b_{\mu}$, $a_{[\mu}b_{\nu]}\equiv a_{\mu}b_{\nu}-a_{\nu}b_{\mu}$,
$a_{\{\mu}b_{\nu\}}\equiv a_{\mu}b_{\nu}+a_{\nu}b_{\mu}$,
$g_{\mu\nu}=\mathrm{diag}(+,-,-,-)$,
$\epsilon^{0123}=-\epsilon_{0123}=1$, and summation over repeated
indices is implied if not stated explicitly. Natural units are
chosen, $c=\varepsilon_0 = \mu_0 = k_{B}=1$, but the reduced Planck constant $\hbar$
is shown explicitly in order to perform the power counting.

\section{Fermions in the CTP formalism}

\label{sec:qed-in-ctp}

\subsection{Lagrangian}

In our units, Planck's constant has dimension [energy $\times $ length].
Not setting $\hbar$ equal to one, as one usually does in natural units,
entails that units of energy are not identical to units of inverse length,
which, in turn, introduces certain ambiguities in the dimensions
of fields and coupling constants, which need to be removed by making additional
definitions for the action of the system. In order to make this explicit for our case,
consider the action of a system of spin-1/2 fermions with mass $m$ interacting
with scalar bosons with mass $M$ via a Yukawa coupling,
\begin{equation}
S [\overline{\psi},\psi,\phi] = \int d^4x \,\mathcal{L}=
\int d^4 x \left[ \overline{\psi}(x) \left(i \hbar \gamma \cdot \partial_x-m\right)\psi(x)
+ g \,\overline{\psi}(x) \phi(x)\psi(x)
- \frac{1}{2} \phi(x) \left( \hbar^2 \partial_x^2 + M^2 \right) \phi(x) \right] \; .
\label{eq:lagrangian_FB}
\end{equation}
Since the action has the
same dimension as $\hbar$, we deduce that the fermion fields
$\psi, \overline{\psi}$ have dimension [length]${}^{-3/2}$, while
the boson field $\phi$ has dimension [energy $\times$ length${}^3$]${}^{-1/2}$.
The dimension of the Yukawa coupling $g$ is [energy $\times$ length]${}^{3/2}$.
Note that, while our unit convention differs from that of
Ref.\ \cite{Itzykson:1980rh}, it is still true that, in momentum space, each propagator
comes with a factor of $\hbar$, while each boson-fermion vertex comes with a
factor of $\hbar^{-1}$.

Integrating out the boson field, we arrive at an action where a scalar fermion current,
$\overline{\psi}(x) \psi(x)$, interacts with another current, $\overline{\psi}(y) \psi(y)$, via
one-boson exchange mediated by the boson propagator $\Delta(x,y)$.
Such a model can be readily generalized to incorporate one-boson exchange
interactions with other quantum numbers,
\begin{equation}
S [\overline{\psi},\psi] = \int d^4 x \,\overline{\psi}(x)
\left(i \hbar \gamma \cdot \partial_x-m\right)\psi(x)
+ \sum_{c} \frac{g_c^2}{\hbar} \int d^4x d^4y \,
\overline{\psi}(x)\Gamma^{(c)}_a \psi(x) \, \Delta^{(c)}_{ab} (x,y)\,
\overline{\psi}(y) \Gamma^{(c)}_{b}\psi(y)\; ,
\label{eq:lagrangian_0}
\end{equation}
where $\Delta^{(c)}_{ab}(x,y)$ is the propagator of the boson of type $c$
(which can be scalar, pseudoscalar, vector,
axial-vector, tensor, etc.\ in space-time, color space, flavor space etc.),
$\Gamma^{(c)}_{a}, \Gamma^{(c)}_b$ are certain products of Dirac, color, and flavor
matrices, properly chosen to reflect the coupling of the boson to the fermions, and
$g_c$ denotes the coupling constant for the coupling of fermions to bosons of type $c$.

The fermion action (\ref{eq:lagrangian_0}) is generic for fermions in all theories
with Yukawa interactions between fermions and bosons, and emerges naturally
after integrating out the boson degrees of freedom
(when neglecting boson self-interactions). It is also
generic for certain gauge theories, such as QED, and even for
QCD when neglecting gluon self-interactions.

Assuming that the range of the boson-exchange interaction is much smaller
than any other scale in the
problem, we may replace $(g_c^2/\hbar) \Delta^{(c)}_{ab}(x,y)
\simeq G_{c} \, \delta_{ab} \delta^{(4)}(x-y)$, where
the 4-fermion coupling constant $G_c$ has dimension [energy $\times$ length$^3$],
and arrive at the Lagrangian of the time-honored NJL
model \citep{Nambu:1961tp,Nambu:1961fr},
\begin{equation}
\mathcal{L}=
\overline{\psi}(x) \left(i\hbar\gamma \cdot \partial_x-m\right)\psi(x)+ \sum_{c} G_{c}\,
\left[\overline{\psi}(x) \Gamma^{(c)}_{a}\psi(x)\right]^2\; .
\label{eq:lagrangian}
\end{equation}
Note that, in momentum space, the four-fermion vertex also carries
a factor of $\hbar^{-1}$.

\subsection{CTP formalism}

\begin{figure}
\caption{\label{fig:ctp}The closed-time path.}

\begin{center}\includegraphics[scale=0.8]{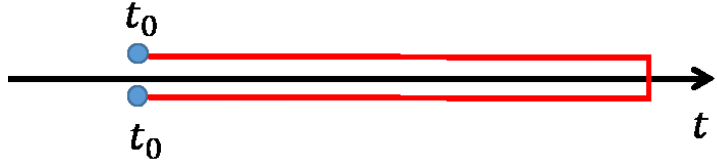}\end{center}
\end{figure}

All information on the non-equilibrium dynamics of the system is
provided by the generating functional for correlation functions on the CTP
\begin{eqnarray}
Z\left[\eta,\overline{\eta},\rho\right] & = & \mathrm{Tr}\left\{ \rho(t_{0})T_{C}
\exp\left[\frac{i}{\hbar}\int_{C}d^{4}x\left(\mathcal{L}
+\overline{\psi}\eta+\overline{\eta}\psi\right)\right]\right\}
\nonumber \\
 & = & \mathrm{Tr}\left\{ \rho(t_{0})T_{C}\exp\left[\frac{i}{\hbar}
 \int_{t_{0}}^{\infty}dt_{+}d^{3}x\left(\mathcal{L}_{+}
 +\overline{\psi}_{+}\eta_{+}+\overline{\eta}_{+}\psi_{+}\right)
 -\frac{i}{\hbar}\int_{t_{0}}^{\infty}dt_{-}
 d^{3}x\left(\mathcal{L}_{-}+\overline{\psi}_{-}\eta_{-}
 +\overline{\eta}_{-}\psi_{-}\right)\right]\right\} \;,
\end{eqnarray}
where $\rho(t_{0})$ is the density matrix at the initial time, $C$
denotes the CTP as shown in Fig.\ \ref{fig:ctp},
$T_{C}$ is the time-ordering operator on the
CTP, and $\eta$ and $\overline{\eta}$ are sources
for $\overline{\psi}$ and $\psi$, respectively. The subscript
$\pm$ denotes quantities on the positive/negative time branch.

The two-point function of the theory can be put into matrix form,
\begin{equation}
G(x_{1},x_{2})=\left(\begin{array}{cc}
G^{++}(x_{1},x_{2}) & G^{+-}(x_{1},x_{2})\\
G^{-+}(x_{1},x_{2}) & G^{--}(x_{1},x_{2})
\end{array}\right)=\left(\begin{array}{cc}
G^{F}(x_{1},x_{2}) & G^{<}(x_{1},x_{2})\\
G^{>}(x_{1},x_{2}) & G^{\bar{F}}(x_{1},x_{2})
\end{array}\right)\; .\label{eq:green-ctp}
\end{equation}
where $G^{ij}(x_{1},x_{2})$ (with $i,j=+,-$) means that the first time argument
$t_{1}=x_{1}^{0}$ lives on the time branch $i$ and the second time argument
$t_{2}=x_{2}^{0}$ lives on the time branch $j$. For $G^{++}$, both $t_{1}$ and $t_{2}$ are
on the positive time branch. Then, the ordering on the CTP is just
the standard time ordering of quantum field theory, so that $G^{++}$
is simply the Feynman propagator $G^{F}$. On the other hand, $G^{-+}$ means
that $t_{1}$ lives on the negative and $t_{2}$ on the positive time branch, respectively,
such that $t_{1}$ is later than $t_{2}$ considering the ordering on the
CTP. Consequently, this two-point function is denoted as $G^{>}$. Analogously,
$G^{+-}\equiv G^{<}$. When both $t_{1}$ and $t_{2}$ live on the negative
time branch, $t_{1}>t_{2}$ actually means that, on the CTP, $t_{2}$ is later
than $t_{1}$, so the ordering on the CTP is actually equivalent
to the standard anti-time ordering, and hence we denote $G^{--}$ as $G^{\bar{F}}$.
The definitions of the various Green's functions are
\begin{eqnarray}
G_{\alpha\beta}^{F}(x_{1},x_{2}) & = &
\left\langle T\psi_{\alpha}(x_{1})\overline{\psi}_{\beta}(x_{2})\right\rangle
\;, \label{eq:def-green_F}\\
G_{\alpha\beta}^{\bar{F}}(x_{1},x_{2}) & = &
\left\langle T_{A}\psi_{\alpha}(x_{1})\overline{\psi}_{\beta}(x_{2})
\right\rangle \; ,\label{eq:def-green_barF} \\
G_{\alpha\beta}^{<}(x_{1},x_{2}) & = &
-\left\langle \overline{\psi}_{\beta}(x_{2})\psi_{\alpha}(x_{1})\right\rangle
\; ,\label{eq:def-green-lesser} \\
G_{\alpha\beta}^{>}(x_{1},x_{2}) & = &
\left\langle \psi_{\alpha}(x_{1})\overline{\psi}_{\beta}(x_{2})\right\rangle
\; ,\label{eq:def-green-larger}
\end{eqnarray}
where $T$ and $T_{A}$ denote the time-ordering and anti-time-ordering
operators, respectively, and angular brackets denote averages weighted
by $\rho(t_{0})$. Not all of the four types of two-point functions appearing in
Eq.\ (\ref{eq:green-ctp}),
or Eqs.\ (\ref{eq:def-green_F}) -- (\ref{eq:def-green-larger}), respectively,
are independent, for example they satisfy
\begin{equation}
G^{F}+G^{\bar{F}}=G^{<}+G^{>}\; ,\label{eq:g-less-g-larger-gff}
\end{equation}
which is a direct consequence of the anticommutation relations for fermion field operators.
Equivalently we can use the following two-point functions, which are linear
combinations of those in Eqs.\ (\ref{eq:def-green_F}) -- (\ref{eq:def-green-larger}),
\begin{eqnarray}
G^{R} & = & G^{F}-G^{<}=-G^{\bar{F}}+G^{>}\; ,\nonumber \\
G^{A} & = & G^{F}-G^{>}=-G^{\bar{F}}+G^{<}\; , \\
G^{C} & = & G^{F}+G^{\bar{F}}=G^{>}+G^{<}\;,\nonumber
\end{eqnarray}
where $G^{R}$ and $G^{A}$ are the retarded and advanced two-point
Green's functions, respectively, the explicit forms of which are given by
\begin{eqnarray}
G^{R} (x_{1},x_{2})& = &
\theta(t_{1}-t_{2})\left[G^{>}(x_{1},x_{2})-G^{<}(x_{1},x_{2})\right] \; ,
\label{eq:gr-explicit} \\
G^{A} (x_{1},x_{2})& = &
-\theta(t_{2}-t_{1})\left[G^{>}(x_{1},x_{2})-G^{<}(x_{1},x_{2})\right]\; .
\label{eq:ga-explicit}
\end{eqnarray}
We can actually express all two-point functions in terms of $G^{>}$ and $G^{<}$
with the help of $\theta(t_{1}-t_{2})$ and $\theta(t_{2}-t_{1})$.
We note that Eqs.\ (\ref{eq:g-less-g-larger-gff}) -- (\ref{eq:ga-explicit})
are also valid for the self-energy $\Sigma(x_1,x_2)$, which is a two-point vertex function.

\section{Kadanoff--Baym equation for fermions}
\label{sec:kb}

\subsection{Kadanoff-Baym equation in quasi-particle approximation}
\label{sec:kbA}

The Kadanoff--Baym (KB) equation \citep{Kadanoff:1962}
can be derived from the Dyson--Schwinger equation on the CTP contour.
For $G^<(x_1,x_2)$, this equation reads \cite{Mrowczynski:1992hq}
\begin{equation} \label{eq:DSE}
\left( i \hbar \gamma \cdot \partial_{x_1} - m \right) G^<(x_1,x_2) =
i \hbar \int d^4x'\left[\Sigma^R(x_1,x')\, G^<(x',x_2)
 + \Sigma^<(x_1,x') G^A (x',x_2) \right]\;.
\end{equation}
The two-point Wigner function is defined by a Fourier transform of the
two-point function (\ref{eq:def-green-lesser})
with respect to the difference $y=x_{1}-x_{2}$ of the two space-time points
$x_1$ and $x_2$,
\begin{equation}
G_{\alpha\beta}^{<}(x,p)\equiv - \int d^{4}y\,e^{ip\cdot y/\hbar}
\left\langle \overline{\psi}_{\beta}\left(x-\frac{y}{2}\right)\psi_{\alpha}\left(x+\frac{y}{2}\right)
\right\rangle\; ,
\label{eq:wigner-def}
\end{equation}
where $x=(x_{1}+x_{2})/2$ is the arithmetic mean (or center) of the
two space-time points $x_1$ and $x_2$.
Up to second order in an expansion
in $\hbar$, the KB equation for the Wigner function $G^{<}(x,p)$
reads \cite{Mrowczynski:1992hq}
\begin{eqnarray}
\left(\gamma \cdot K -m\right)G^{<}(x,p)
& = & i\hbar\left[\Sigma^{R}(x,p)G^{<}(x,p)+\Sigma^{<}(x,p)G^{A}(x,p)\right] \nonumber \\
&   & + \, \frac{\hbar^{2}}{2}\left[\left\{ \Sigma^{R}(x,p),G^{<}(x,p)\right\} _{\mathrm{PB}}
+\left\{ \Sigma^{<}(x,p),G^{A}(x,p)\right\} _{\mathrm{PB}}\right]\;,\label{eq:kb-1}
\end{eqnarray}
where we defined the operator
\begin{equation}
K^\mu \equiv p^\mu + \frac{i\hbar}{2} \partial_{x}^\mu\;,\label{eq:k-op}
\end{equation}
and where $\{A,B\}_{\mathrm{PB}}$ denotes the Poisson bracket
\begin{equation}
\left\{ A,B\right\} _{\mathrm{PB}}\equiv (\partial_{x} A) \cdot (\partial_{p}B)
-(\partial_{p}A)\cdot (\partial_{x}B)\; .
\end{equation}
Note that $G^{i}(x,p), \, i = <,>,R,A$ is formally of order $\hbar$ [cf.\ remarks after
Eq.\ (\ref{eq:lagrangian_FB})].
We will tacitly factor out one power of $\hbar$ from the $G^{i}(x,p)$ on both sides
of Eq.\ (\ref{eq:kb-1}), such that an expansion of $G^{i}(x,p)$ in powers of $\hbar$
starts at order $\hbar^0$.

We will work to lowest diagrammatic order
in the $T$-matrix approximation for the self-energies
$\Sigma^{i}, \, i = <,>,R,A$. The $T$-matrix approximation
is appropriate for the description of binary collisions, because to lowest order in
this approximation,
the self-energy contains two loops, so that upon cutting the respective
diagram, one obtains two on-shell particles in the in- and the outgoing channel.
Such a two-loop diagram is of order $\mathcal{O}(G_c^2)$
in the coupling constant of the NJL Lagrangian (\ref{eq:lagrangian}). We
will neglect all diagrams of higher order in loops, or equivalently, of higher order
in the coupling constant $G_c$.
Following the standard Feynman rules in momentum space,
where each propagator comes with a factor of $\hbar$ and a
four-fermion vertex comes with
a factor of $\hbar^{-1}$, such a diagram is of order $\hbar$.
This factor is explicitly pulled out from $\Sigma^{i}$ in Eq.\ (\ref{eq:kb-1}), such
that the $\hbar$ expansion of $\Sigma^{i}$ starts at order $\hbar^0$,
just as that of $G^{i}$. The second term
in Eq.\ (\ref{eq:kb-1}) has an extra factor of $\hbar$,
since the Poisson bracket involves the product
of a spatial and a momentum derivative.

Equation (\ref{eq:kb-1}) is consistent with Ref.\ \citep{Schonhofen:1994zf},
except for factors of $i$, since we absorbed an $i$ into the definition of
the two-point Green's functions in Eqs.\
(\ref{eq:def-green_F}) -- (\ref{eq:def-green-larger}).
The adjoint KB equation can be obtained from
Eq.\ (\ref{eq:kb-1}) by exchanging $\Sigma\leftrightarrow G$
on the right-hand side and by pulling $G^{<}$ to the left of the
differential operator, changing $\partial_{x}\rightarrow-\overleftarrow{\partial}_{x}$
on the left-hand side,
\begin{eqnarray}
 G^{<}(x,p)\left( \gamma \cdot \overleftarrow{K}^*-m\right)
 & = & i\hbar\left[G^{R}(x,p)\Sigma^{<}(x,p)+G^{<}(x,p)\Sigma^{A}(x,p)\right] \nonumber \\
&  & + \, \frac{\hbar^{2}}{2}\left[\left\{ G^{R}(x,p),\Sigma^{<}(x,p)\right\} _{\mathrm{PB}}
 +\left\{ G^{<}(x,p),\Sigma^{A}(x,p)\right\} _{\mathrm{PB}}\right]\;.\label{eq:kb-2}
\end{eqnarray}
Similarly one can derive the KB equations for
$G^{>}$ from Eqs.\ (\ref{eq:kb-1}), (\ref{eq:kb-2})
by replacing $G^{<}\rightarrow G^{>}$ and $\Sigma ^{<}\rightarrow \Sigma ^{>}$ but
keeping all retarded and advanced quantities unchanged.

Taking the Fourier transform
of Eqs.\ (\ref{eq:gr-explicit}), (\ref{eq:ga-explicit})
and the equivalent relation for $\Sigma$,
the retarded and advanced two-point (vertex) function in momentum
space can be expressed as
\begin{eqnarray}
O^{R/A}(x,p) & = & \frac{1}{2\pi i}\int dk_{0}\frac{1}{k_{0}-p_{0}\mp i\epsilon}
\left[O^{>}(x,k_{0},\mathbf{p})-O^{<}(x,k_{0},\mathbf{p})\right]\nonumber \\
 & = & \pm\frac{1}{2}\left[O^{>}(x,p)-O^{<}(x,p)\right]+\frac{1}{2\pi i}\mathscr{P}
 \int dk_{0}\frac{1}{k_{0}-p_{0}}\left[O^{>}(x,k_{0},\mathbf{p})
 -O^{<}(x,k_{0},\mathbf{p})\right]\;,
 \label{eq:retarded-advanced}
\end{eqnarray}
where $O=\Sigma$ or $G$. The principal
value of the integral is denoted by $\mathscr{P}$. The second term represents
off-shell contributions, including either off-mass-shell or energy
non-conservation effects.
In the quasiparticle approximation, we can
neglect the principal-value part of Eq.\ (\ref{eq:retarded-advanced}),
and Eq.\ (\ref{eq:kb-1}) can be put into the form
\begin{equation}
\left(\gamma \cdot K-m\right)G^{<}(x,p) = I_{\mathrm{coll}}\,,
\label{eq:kb-main}
\end{equation}
where the collision term $I_{\mathrm{coll}}$ is given by
\begin{eqnarray}
I_{\mathrm{coll}}& \equiv & -\frac{i \hbar}{2}\left[\Sigma^{<}(x,p)G^{>}(x,p)
-\Sigma^{>}(x,p)G^{<}(x,p)\right]
\nonumber \\
 &  & -\frac{\hbar^{2}}{4}\left[\left\{ \Sigma^{<}(x,p),G^{>}(x,p)\right\} _{\mathrm{PB}}
 -\left\{ \Sigma^{>}(x,p),G^{<}(x,p)\right\} _{\mathrm{PB}}\right]\;.
\label{eq:collision_term}
\end{eqnarray}
The above equation is our starting point for deriving the Boltzmann
equation for spin-1/2 particles, with a collision term including contributions
up to next-to-leading order in $\hbar$. Similarly one can derive the KB equation
for $G^{>}$ from Eq.\ (\ref{eq:kb-main}) by replacing $G^{<}\rightarrow G^{>}$
on the left-hand side, while
the collision term remains unchanged.

\subsection{Clifford decomposition}

We can expand $G^{<}(p,X)$ in terms of the 16 independent generators
of the Clifford algebra, $\Gamma_a$, $a = 1, \ldots,16$, with
\begin{equation}
\Gamma_a \in \left\{ 1,\, \gamma^{\mu}, \, i\gamma^{5}
=-\gamma^{0}\gamma^{1}\gamma^{2}\gamma^{3}, \,
\gamma^{5}\gamma^{\mu},\, \sigma^{\mu\nu}
=\frac{i}{2}\left[\gamma^{\mu},\gamma^{\nu}\right]
\right\}\;, \label{eq:c-generator}
\end{equation}
such that
\begin{equation}
G^{<}(x,p)=\frac{1}{4}\left(\mathcal{F}+i\gamma^{5}\mathcal{P}
+\gamma^\mu \mathcal{V}_{\mu}
+\gamma^{5}\gamma^{\mu}\mathcal{A}_{\mu}
+\frac{1}{2}\sigma^{\mu\nu}\mathcal{S}_{\mu\nu}\right)\;.
\label{eq:wigner-decomp}
\end{equation}
The real-valued coefficient functions $\mathcal{F}$, $\mathcal{P}$, $\mathcal{V}_{\mu}$,
$\mathcal{A}_{\mu}$, and $\mathcal{S}_{\mu\nu}$ are the scalar, pseudo-scalar,
vector, axial-vector, and tensor components of $G^{<}(x,p)$, respectively, which can
be obtained by taking the trace of $G^{<}(x,p)$ multiplied with the appropriate
generator $\Gamma_a$ of the Clifford algebra. Note the sign convention of
$\mathcal{A}^{\mu}=\mathrm{Tr}\left(\gamma^{\mu}\gamma^{5}G^{<}\right)$,
which is the same as in Ref.\ \citep{Gao:2012ix,Weickgenannt:2019dks},
but different from Ref.\ \citep{Vasak:1987um}.

We can rewrite Eq.\ (\ref{eq:kb-main}) by inserting Eq.\ (\ref{eq:wigner-decomp})
into the left-hand side, multiplying with the appropriate $\Gamma_a$, and taking the trace.
Then we obtain a system of coupled equations for the components
of the Wigner function. The real parts of these equations read
\begin{eqnarray}
p^\mu \mathcal{V}_\mu - m \mathcal{F}
& = & \mathrm{Re\, Tr} \left( I_{\mathrm{coll}}\right)\;,
\label{eq:scalar_real} \\
m \mathcal{P} + \frac{\hbar}{2}\, \partial_x^\mu \mathcal{A}_\mu
& = &  \mathrm{Re\, Tr} \left(i \gamma^5 I_{\mathrm{coll}}\right)\;,
\label{eq:pseudoscalar_real}  \\
p_\mu \mathcal{F} - m \mathcal{V}_\mu
+ \frac{\hbar}{2} \, \partial_x^\nu \mathcal{S}_{\mu \nu}
& = & \mathrm{Re\, Tr} \left( \gamma_\mu I_{\mathrm{coll}}\right)\;,
\label{eq:vector_real}  \\
\frac{1}{2} \, \epsilon_{\mu \nu \alpha \beta} p^\nu \mathcal{S}^{\alpha \beta}
+ m \mathcal{A}_\mu
- \frac{\hbar}{2} \, \partial_{x,\mu} \mathcal{P} & = &
\mathrm{Re\, Tr} \left( \gamma^5 \gamma_\mu I_{\mathrm{coll}}\right)\;,
\label{eq:axialvector_real}  \\
\epsilon_{\mu \nu \alpha \beta} p^\alpha \mathcal{A}^\beta + m \mathcal{S}_{\mu \nu}
- \frac{\hbar}{2} \partial_{x[\mu} \mathcal{V}_{\nu]}
& = & - \mathrm{Re\, Tr} \left( \sigma_{\mu \nu} I_{\mathrm{coll}}\right)\;,
\label{eq:tensor_real}
\end{eqnarray}
while the imaginary parts are
\begin{eqnarray}
\frac{\hbar}{2}\, \partial_x^\mu \mathcal{V}_\mu & = &
\mathrm{Im\, Tr} \left( I_{\mathrm{coll}}\right)\;,
\label{eq:scalar_im} \\
p^\mu \mathcal{A}_\mu & = & \mathrm{Im\, Tr} \left(- i \gamma^5 I_{\mathrm{coll}}\right)\;,
\label{eq:pseudoscalar_im} \\
p^\nu \mathcal{S}_{\nu \mu} + \frac{\hbar}{2} \, \partial_{x,\mu} \mathcal{F}
& = & \mathrm{Im\, Tr} \left( \gamma_\mu I_{\mathrm{coll}}\right)\;, \label{eq:vector_im} \\
p_\mu \mathcal{P} + \frac{\hbar}{4} \, \epsilon_{\mu \nu \alpha \beta}
\partial_x^\nu \mathcal{S}^{\alpha \beta}
& = & \mathrm{Im\, Tr} \left( \gamma^5 \gamma_\mu I_{\mathrm{coll}}\right)\;,
\label{eq:axialvector_im} \\
p_{[\mu} \mathcal{V}_{\nu]} + \frac{\hbar}{2}\, \epsilon_{\mu \nu \alpha \beta}
\partial_x^\alpha
\mathcal{A}^\beta & = & - \mathrm{Im\, Tr} \left( \sigma_{\mu \nu} I_{\mathrm{coll}}\right)\;,
\label{eq:tensor_im}
\end{eqnarray}
In the absence of collisions, Eqs.\ (\ref{eq:scalar_real}) -- (\ref{eq:tensor_im})
are equivalent to Eqs.\ (11) -- (20) of Ref.\
\cite{Weickgenannt:2019dks} without electromagnetic fields.
Equations (\ref{eq:pseudoscalar_real}), (\ref{eq:vector_real}),
(\ref{eq:tensor_real}), (\ref{eq:scalar_im}),
(\ref{eq:pseudoscalar_im}), and (\ref{eq:tensor_im}) correspond to
Eqs.\ (6) -- (11) of Ref.\ \cite{Weickgenannt:2020aaf}, where the collision term
was derived using a different method.
In the following, we will write Eqs.\ (\ref{eq:scalar_real}) --
(\ref{eq:tensor_im}) perturbatively order by order in $\hbar$.

There is an equivalent way of expressing the Dirac-like KB equation (\ref{eq:kb-main})
in terms of components of the Wigner function. Acting with
the operator $\gamma\cdot K + m$ onto Eq.\ (\ref{eq:kb-main}) and combining
the resulting equation with its Hermitian conjugate, multiplied from
the left- and the right-hand side with $\gamma^0$, we can
derive a Klein-Gordon-type equation, which gives an on-shell condition,
and a Boltzmann-type equation,
\begin{eqnarray}
\left(p^{2}-\frac{\hbar^{2}}{4}\partial_x^{2}-m^{2}\right)G^{<}(x,p)
& = & \frac{1}{2}\left\{(\gamma \cdot K+m)I_{\mathrm{coll}}+\gamma^{0}
\left[(\gamma \cdot K+m)I_{\mathrm{coll}}\right]^{\dagger}\gamma^{0}\right\}\,;\\
\label{eq:massshell-wig}
\hbar\, p\cdot\partial_x G^{<}(x,p) & = &
-\frac{i}{2}\left\{ (\gamma \cdot K+m)I_{\mathrm{coll}}-\gamma^{0}
\left[(\gamma \cdot K+m)I_{\mathrm{coll}}\right]^{\dagger}\gamma^{0}\right\} \;,
\label{eq:boltzmann-wig}
\end{eqnarray}
where we have used
$\gamma^0 (G^{<})^\dagger \gamma^0 \equiv G^<$, which can be derived from
Eq.\ (\ref{eq:wigner-decomp}).
Taking the trace with the appropriate basis elements of
the Clifford decomposition, we obtain for the components of the Wigner function
\begin{eqnarray}
\left(p^{2} -\frac{\hbar^{2}}{4}\partial_x^{2}-m^{2}\right)\,
\mathrm{Tr}\left(\Gamma_{a}G^{<}\right) & = &
\mathrm{Re}\mathrm{Tr}\left[\Gamma_{a}(\gamma\cdot K+m)I_{\mathrm{coll}}\right]\;,
\label{eq:massshell-wigner} \\
\hbar\,  p\cdot\partial_x\, \mathrm{Tr}\left(\Gamma_{a}G^{<}\right) & = &
\mathrm{Im}\mathrm{Tr}\left[\Gamma_{a}(\gamma\cdot K+m)I_{\mathrm{coll}}\right] \;.
\label{eq:boltzmann-wigner}
\end{eqnarray}
Equations (\ref{eq:massshell-wigner}) and (\ref{eq:boltzmann-wigner}) are
equivalent to  Eqs.\ (\ref{eq:scalar_real}) -- (\ref{eq:tensor_im}).

\subsection{Cancellation of off-shell terms}
\label{sec:cancel-off}

We now show that off-shell terms cancel in the Boltzmann-type equations
(\ref{eq:boltzmann-wigner}) to lowest order in the coupling constant. We
can decompose the Wigner function into an on-shell and off-shell part as
\begin{equation}
G^{<}(x,p)=G_{\mathrm{on}}^{<}(x,p)+G_{\mathrm{off}}^{<}(x,p)\;,
\end{equation}
where the on-shell part fulfills
$(p^2 - m^2)G_{\mathrm{on}}^{<} (x,p) \equiv 0$ up to order $\mathcal{O}(\hbar)$.
Thus we obtain from the on-shell condition (\ref{eq:massshell-wigner})
\begin{equation}
\mathrm{Tr}\left(\Gamma_{a}G_{\mathrm{off}}^{<}\right)
= \frac{1}{p^{2}-m^{2}}\,
\mathrm{Re}\mathrm{Tr}\left[\Gamma_{a}(\gamma\cdot K+m)I_{\mathrm{coll}}\right]
+ \mathcal{O}(\hbar^2)\;.
\end{equation}
Inserting this into the Boltzmann equation (\ref{eq:boltzmann-wigner}), we obtain
\begin{eqnarray}
\hbar\, p\cdot\partial_x\mathrm{Tr}\left(\Gamma_{a}G_{\mathrm{on}}^{<}\right)
& = & \mathrm{Im}\mathrm{Tr}\left[\Gamma_{a}(\gamma\cdot K+m)I_{\mathrm{coll}}\right]
 -\frac{\hbar}{p^{2}-m^{2}}\,p\cdot\partial_x\,
 \mathrm{Re}\mathrm{Tr}\left[\Gamma_{a}(\gamma\cdot K+m)I_{\mathrm{coll}}\right]
 + \mathcal{O}(\hbar^3)\;.
 \label{eq:on-shell-boltzmann}
\end{eqnarray}
We can also decompose the collision term into an on-shell and off-shell part as
\begin{equation}
I_{\mathrm{coll}}=I_{\mathrm{coll}}^{(\mathrm{on})}+I_{\mathrm{coll}}^{(\mathrm{off})}\;,
\label{coll-on-off}
\end{equation}
such that Eq.\ (\ref{eq:on-shell-boltzmann}) becomes
\begin{equation}
\hbar \, p\cdot\partial_x\, \mathrm{Tr}\left(\Gamma_{a}G_{\mathrm{on}}^{<}\right)
 =  \mathrm{Im}\mathrm{Tr}\left[\Gamma_{a}(\gamma\cdot K+m)
I_{\mathrm{coll}}^{(\mathrm{on})}\right]
+\mathrm{Im}\mathrm{Tr}\left[\Gamma_{a}(\gamma\cdot K+m)
\left(I_{\mathrm{coll}}^{(\mathrm{off})}-i\frac{\hbar}{p^{2}-m^{2}}\,p\cdot\partial_x \,
 I_{\mathrm{coll}}\right)\right]+ \mathcal{O}(\hbar^3)\;.
 \label{eq:on-shell-boltz}
\end{equation}
From Eqs.\ (\ref{eq:kb-2}) and (\ref{eq:collision_term}) we conclude that the
collision term satisfies
\begin{equation}
I_{\mathrm{coll}}\left(\gamma\cdot p-m
-\frac{i\hbar}{2}\gamma\cdot\overleftarrow{\partial}_x\right)
 =\mathcal{O}(\hbar^2 G_c^{4})\;.
\label{another-eq-coll}
\end{equation}
This is of higher order in the coupling constant and will thus be set to zero in the
following.
Note that the derivative $\partial_x$ acts only on $G^{\lessgtr}$ but not on
$\Sigma^{\lessgtr}$ in $I_{\mathrm{coll}}$, cf.\ Eq.\ (\ref{eq:collision_term}), since it comes
from $\partial_{x_2}$ before Fourier transformation of $y=x_1-x_2$.
We act with the operator $\left(\gamma\cdot p+m
-\frac{i\hbar}{2}\gamma\cdot\overleftarrow{\partial}_x\right)$
from the right-hand side onto the above equation to obtain
\begin{equation}
I_{\mathrm{coll}}\left[\left(p^{2}-
\frac{\hbar^{2}}{4}\overleftarrow{\partial}_x^{2}-m^{2}\right)
-i\hbar \, p\cdot\overleftarrow{\partial}_x\right]=\mathcal{O}(\hbar ^2 G_c^{4})\;.
\label{i-coll-eom}
\end{equation}
With Eq.\ (\ref{coll-on-off}), Eq.\ (\ref{i-coll-eom}) leads to
\begin{equation}
I_{\mathrm{coll}}^{(\mathrm{off})}= i\frac{\hbar}{p^{2}-m^{2}}\,
p\cdot\partial_x\,I_{\mathrm{coll}} + \mathcal{O}(\hbar ^2 G_c^{4})\;.
\end{equation}
Thus, Eq.\ (\ref{eq:on-shell-boltz}) leads to
Boltzmann equations for the on-shell components of the Wigner function
\begin{equation}
\hbar \, p\cdot\partial_x\, \mathrm{Tr}\left(\Gamma_{a}G_{\mathrm{on}}^{<}\right)
=\mathrm{Im}\mathrm{Tr}\left[\Gamma_{a}(\gamma\cdot K+m)
I_{\mathrm{coll}}^{(\mathrm{on})}\right]+\mathcal{O}(\hbar ^2 G_c^{4})\;.
\label{onshell-boltzmann}
\end{equation}
This concludes the proof that, to lowest order in the coupling
constant, off-shell contributions cancel in the Boltzmann equation. A similar
result using a different method has been obtained in
Refs.\ \cite{Weickgenannt:2020aaf,Weickgenannt:2021cuo}.

\section{Semiclassical expansion of KB equation}

\label{sec:h-expand-kb}
In this section we derive the system of equations
(\ref{eq:scalar_real}) -- (\ref{eq:tensor_im})
order by order in powers of $\hbar$. To this end,
we expand every quantity (functions as well as operators) as
\begin{equation} \label{eq:hbar_expansion}
F=\sum_{n=0}^{\infty}\hbar^{n}F^{(n)}\;,
\end{equation}
and truncate this expansion at a given order $n$.

\subsection{Zeroth order in $\hbar$}
\label{zeroth-order-h-expand}

At $\mathcal{O}(\hbar^{0})$, the collision term (\ref{eq:collision_term}) vanishes,
$I_{\mathrm{coll}}^{(0)} =0$,
since it is at least of order $\mathcal{O}(\hbar)$.
Equations (\ref{eq:scalar_real}) -- (\ref{eq:tensor_im}) become
\begin{eqnarray}
p^{\mu}\mathcal{V}_{\mu}^{(0)}-m\mathcal{F}^{(0)} & = & 0\;,\label{eq:scalar_real_0} \\
\mathcal{P}^{(0)} & = & 0\;,\label{eq:pseudoscalar_real_0} \\
p_{\mu}\mathcal{F}^{(0)}-m\mathcal{V}_{\mu}^{(0)}& = & 0\;,\label{eq:vector_real_0}\\
\frac{1}{2}\, \epsilon_{\mu\nu \alpha\beta}p^\nu\mathcal{S}^{(0)\alpha\beta}
+m\mathcal{A}_{\mu}^{(0)} & = & 0\;,\label{eq:axialvector_real_0} \\
\epsilon_{\mu \nu \alpha \beta} p^\alpha \mathcal{A}^{(0)\beta}
+ m \mathcal{S}^{(0)}_{\mu \nu}
& = & 0\;, \label{eq:tensor_real_0}
\end{eqnarray}
and
\begin{eqnarray}
p^{\mu}\mathcal{A}^{(0)}_{\mu}& = & 0\;,\label{eq:pseudoscalar_im_0} \\
p^{\nu}\mathcal{S}^{(0)}_{\mu \nu} & = & 0\;,\label{eq:vector_im_0} \\
p_{[\mu}\mathcal{V}^{(0)}_{\nu]}& = & 0\;, \label{eq:tensor_im_0}
\end{eqnarray}
respectively. In the following, we choose the scalar and axial-vector components
$\mathcal{F}$ and $\mathcal{A}_{\mu}$
as independent ones. Then
we can express the other components in terms of
$\mathcal{F}$ and $\mathcal{A}_{\mu}$ using
Eqs.\ (\ref{eq:pseudoscalar_real_0}),
(\ref{eq:vector_real_0}), and (\ref{eq:tensor_real_0}), respectively,
\begin{eqnarray}
\mathcal{P}^{(0)} & = & 0\;,\label{eq:pseudoscalar_real_0a} \\
\mathcal{V}_{\mu}^{(0)}& = & \frac{1}{m} p_{\mu}\mathcal{F}^{(0)}\;,
\label{eq:vector_real_0a}\\
\mathcal{S}^{(0)}_{\mu \nu}& = &
-\frac{1}{m}\epsilon_{\mu \nu \alpha \beta} p^\alpha \mathcal{A}^{(0)\beta}\;.
\label{eq:tensor_real_0a}
\end{eqnarray}
The solutions for all components at $\mathcal{O}(\hbar^{0})$ are known and given
in Eq.\ (28) of Ref.\ \citep{Weickgenannt:2019dks}.
Following Eqs.\ (\ref{eq:massshell-wigner})
and (\ref{eq:boltzmann-wigner}), all components at $\mathcal{O}(\hbar^{0})$ satisfy
on-shell conditions and kinetic equations.

\subsection{First order in $\hbar$}

At $\mathcal{O}(\hbar)$,
the collision term (\ref{eq:collision_term}) is given by
\begin{equation}
I_{\mathrm{coll}}^{(1)}=-\frac{i}{2} \left[\Sigma^{<(0)}(x,p)G^{>(0)}(x,p)
-\Sigma^{>(0)}(x,p)G^{<(0)}(x,p)\right]\;,
\label{eq:collision-1st}
\end{equation}
where all propagators appearing in $\Sigma^{\gtrless(0)}(x,p)$ are
taken at zeroth order in $\hbar$. We note that the Wigner functions
$G^{\gtrless(0)}(x,p)$ are on-shell
since they satisfy Eq.\ (\ref{eq:massshell-wig}) with $I_{\mathrm{coll}}^{(0)}=0$,
so $I_{\mathrm{coll}}^{(1)}$ in Eq.\ (\ref{eq:collision-1st}) is also on-shell.

Equations (\ref{eq:scalar_real}) -- (\ref{eq:tensor_real}) read to first order in $\hbar$:
\begin{eqnarray}
p^{\mu}\mathcal{V}_{\mu}^{(1)}-m\mathcal{F}^{(1)}
& = & \mathrm{Re\, Tr}\left(I_{\mathrm{coll}}^{(1)}\right)\;,\label{eq:scalar_real_1} \\
\frac{1}{2}\, \partial^{\mu}_{x}\mathcal{A}^{(0)}_{\mu}+m\mathcal{P}^{(1)}
& = & \mathrm{Re\, Tr}\left(i\gamma^{5}I_{\mathrm{coll}}^{(1)}\right)\;,
\label{eq:pseudoscalar_real_1} \\
\frac{1}{2}\, \partial_{x}^{\nu}\mathcal{S}_{\nu\mu}^{(0)}- p_{\mu}\mathcal{F}^{(1)}
+ m\mathcal{V}_{\mu}^{(1)}
& = & - \mathrm{Re\, Tr}\left(\gamma_{\mu}I_{\mathrm{coll}}^{(1)}\right)\;,
\label{eq:vector_real_1} \\
\frac{1}{2}\, \epsilon_{\mu\nu\alpha\beta}p^{\nu}
\mathcal{S}^{(1)\alpha\beta} +m\mathcal{A}_{\mu}^{(1)}
& = & \mathrm{Re\,Tr}\left(\gamma^{5}\gamma_{\mu}I_{\mathrm{coll}}^{(1)}\right)\;,
\label{eq:axialvector_real_1} \\
\frac{1}{2}\,\partial_{x[\mu}\mathcal{V}_{\nu]}^{(0)}
-\epsilon_{\mu\nu\alpha\beta}p^{\alpha}\mathcal{A}^{(1)\beta}
-m\mathcal{S}_{\mu\nu}^{(1)}
& = & \mathrm{Re\, Tr}\left(\sigma_{\mu\nu}I_{\mathrm{coll}}^{(1)}\right)\;,
\label{eq:tensor_real_1}
\end{eqnarray}
where we have used Eq.\ (\ref{eq:pseudoscalar_real_0}) to simplify
Eq.\ (\ref{eq:axialvector_real_1}).
On the other hand, Eqs.\ (\ref{eq:scalar_im}) -- (\ref{eq:tensor_im}) read to first order in
$\hbar$:
\begin{eqnarray}
\frac{1}{2}\, \partial^{\mu}_{x}\mathcal{V}^{(0)}_{\mu}
& = & \mathrm{Im\, Tr}\left(I_{\mathrm{coll}}^{(1)}\right)\;,\label{eq:scalar_im_1} \\
p^\mu \mathcal{A}^{(1)}_\mu
& = & \mathrm{Im\, Tr}\left(-i \gamma^{5}I_{\mathrm{coll}}^{(1)}\right)\;,
\label{eq:pseudoscalar_im_1} \\
\frac{1}{2}\,\partial_{x,\mu}\mathcal{F}^{(0)}+ p^{\nu}\mathcal{S}_{\nu\mu}^{(1)}
& = & \mathrm{Im\, Tr}\left(\gamma_{\mu}I_{\mathrm{coll}}^{(1)}\right)\;,
\label{eq:vector_im_1} \\
\frac{1}{4}\, \epsilon_{\mu\nu\alpha\beta} \partial_{x}^{\nu}\mathcal{S}^{(0)\alpha\beta}
+ p_{\mu}\mathcal{P}^{(1)}
& = & \mathrm{Im \, Tr}\left(\gamma^{5}\gamma_{\mu}I_{\mathrm{coll}}^{(1)}\right)\;,
\label{eq:axialvector_im_1} \\
\frac{1}{2}\,\epsilon_{\mu\nu\alpha\beta}\partial_{x}^{\alpha}
\mathcal{A}^{(0)\beta} + p_{[\mu}\mathcal{V}_{\nu]}^{(1)}
& = & -\mathrm{Im\,Tr}\left(\sigma_{\mu\nu}I_{\mathrm{coll}}^{(1)}\right)\;.
\label{eq:tensor_im_1}
\end{eqnarray}

Contracting Eq.\ (\ref{eq:vector_real_0}) with $\partial^\mu_x$ and then using
Eq.\ (\ref{eq:scalar_im_1}), we derive the Boltzmann equation for the
zeroth-order scalar component at $\mathcal{O}(\hbar)$,
\begin{equation}
p\cdot\partial_{x}\mathcal{F}^{(0)}
=  2m\,  \mathrm{Im\, Tr}\left( I_{\mathrm{coll}}^{(1)}\right)
\;.\label{eq:boltzmann_scalar-1st}
\end{equation}
Alternatively, we can derive this equation by contracting Eq.\ (\ref{eq:vector_im_1})
with $p^\mu$ and using the fact that $I_{\mathrm{coll}}^{(1)}$
is on-shell, cf.\ discussion after Eq.\ (\ref{eq:collision-1st}), which allows to replace
$p \cdot  \gamma \rightarrow m$ under the trace.

On the other hand, multiplying Eq.\ (\ref{eq:tensor_im_1}) with
$p_\sigma\epsilon^{\rho\sigma \mu \nu}$
and using Eq.\ (\ref{eq:pseudoscalar_im_0}),
we derive the Boltzmann equation for the
zeroth-order axial-vector component at $\mathcal{O}(\hbar)$,
\begin{equation}
p\cdot\partial_{x}\mathcal{A}^{(0)\mu}
 = -\epsilon^{\mu\nu\alpha \beta} p_{\nu}\,
\mathrm{Im\,Tr}\left(\sigma_{\alpha\beta}I_{\mathrm{coll}}^{(1)}\right)\;.
\label{eq:boltzmann_axialvector-1st}
\end{equation}
Another way of deriving this equation is by multiplying Eq.\ (\ref{eq:tensor_real_0})
with $\epsilon^{\mu \nu \rho \sigma} \partial_{x,\sigma}$ and using
Eqs.\ (\ref{eq:pseudoscalar_real_1})
and (\ref{eq:axialvector_im_1}), as well as the well-known identity
$\gamma_5 \sigma^{\mu \nu} = \frac{i}{2} \epsilon^{\mu \nu \alpha \beta}
\sigma_{\alpha \beta}$ and using the fact that $I_{\mathrm{coll}}^{(1)}$
is on-shell, which allows to replace
$\mathrm{Tr}\left( p \cdot  \gamma \gamma_5 I_{\mathrm{coll}}^{(1)}
\right) \rightarrow m\, \mathrm{Tr}\left( \gamma_5 I_{\mathrm{coll}}^{(1)}
\right)$.

Multiplying Eq.\ (\ref{eq:scalar_real_1}) with $m$,
Eq.\ (\ref{eq:vector_real_1}) with $p^\mu$,
subtracting both equations, and using Eq.\ (\ref{eq:vector_im_0})
as well as the on-shell condition for $I_{\mathrm{coll}}^{(1)}$,
we derive a mass-shell constraint for $\mathcal{F}^{(1)}$,
\begin{equation} \label{eq:mass-shell_f1}
(p^2-m^2) \mathcal{F}^{(1)} = 2 m \, \mathrm{Re\, Tr} \left(I_{\mathrm{coll}}^{(1)}\right)\;.
\end{equation}
The mass-shell constraint for $\mathcal{A}^{(1)}_\mu$ follows from
Eqs.\ (\ref{eq:axialvector_real_1}) and (\ref{eq:tensor_real_1}),
using Eqs.\ (\ref{eq:tensor_im_0}) and (\ref{eq:pseudoscalar_im_1}),
as well as the on-shell condition for $I_{\mathrm{coll}}^{(1)}$,
\begin{equation} \label{eq:mass-shell_a1}
(p^{2}-m^{2})\mathcal{A}^{(1)}_{\mu}  =  - \epsilon_{\mu\nu\alpha\beta}p^{\nu}\,
\mathrm{Re\, Tr}\left(\sigma^{\alpha\beta}I_{\mathrm{coll}}^{(1)}\right) \;.
\end{equation}
Multiplying Eq.\ (\ref{eq:axialvector_real_1}) with
$p_\lambda \epsilon^{\lambda \rho \sigma \mu}$ and Eq.\ (\ref{eq:tensor_real_1})
with $m$, adding
both equations, and using Eqs.\ (\ref{eq:vector_real_0}), (\ref{eq:vector_im_1}), we obtain
a mass-shell constraint for $\mathcal{S}^{(1)}_{\mu \nu}$,
\begin{equation} \label{eq:mass-shell_s1}
(p^2-m^2) \mathcal{S}^{(1)}_{\mu \nu} =
2 p_{[\mu} \, \mathrm{Im \, Tr} \left( \gamma_{\nu]} I_{\mathrm{coll}}^{(1)} \right)
+ 2m \, \mathrm{Re\, Tr}\left( \sigma_{\mu \nu} I_{\mathrm{coll}}^{(1)}\right) \;,
\end{equation}
where we again used the on-shell condition for $I_{\mathrm{coll}}^{(1)}$.
The mass-shell constraint for $\mathcal{P}^{(1)}$ can be obtained
from Eqs.\ (\ref{eq:pseudoscalar_real_1}) and (\ref{eq:axialvector_im_1}),
using Eq.\ (\ref{eq:axialvector_real_0}) and the on-shell condition for
$I_{\mathrm{coll}}^{(1)}$,
\begin{equation} \label{eq:mass-shell_p1}
(p^{2}-m^{2})\mathcal{P}^{(1)}  =  0\;.
\end{equation}
The mass-shell constraint for $\mathcal{V}^{(1)}_\mu$ can be derived from Eqs.\
(\ref{eq:vector_real_1}) and (\ref{eq:tensor_im_1}), using Eqs.\ (\ref{eq:tensor_real_0})
and (\ref{eq:scalar_real_1}), as well as the on-shell condition
for $I_{\mathrm{coll}}^{(1)}$,
\begin{equation} \label{eq:mass-shell_v1}
(p^{2}-m^{2})\mathcal{V}_{\mu}^{(1)}  =
2 p_{\mu}\, \mathrm{Re \,Tr}\left(I_{\mathrm{coll}}^{(1)}\right)\;.
\end{equation}

Once $\mathcal{F}^{(1)}$ and $\mathcal{A}_{\mu}^{(1)}$ are known, we can
obtain all other components at order $\mathcal{O}(\hbar)$ via
Eqs.\ (\ref{eq:pseudoscalar_real_1}),
(\ref{eq:vector_real_1}), and (\ref{eq:tensor_real_1}),
\begin{eqnarray}
\mathcal{P}^{(1)} & = &-\frac{1}{2m}\partial_x^{\mu}\mathcal{A}^{(0)}_{\mu}
+\frac{1}{m}\mathrm{Re\, Tr}
\left(i \gamma^{5}I_{\mathrm{coll}}^{(1)}\right)\;,\label{eq:sol-p1} \\
\mathcal{V}_{\mu}^{(1)} & = &\frac{1}{m}p_{\mu}\mathcal{F}^{(1)}
-\frac{1}{2m}\partial_{x}^{\nu}\mathcal{S}_{\nu\mu}^{(0)}
-\frac{1}{m}\mathrm{Re \,Tr}\left(\gamma_{\mu} I_{\mathrm{coll}}^{(1)}\right)\; ,
\label{eq:sol-v1} \\
\mathcal{S}_{\mu\nu}^{(1)} &=& -\frac{1}{m}\epsilon_{\mu\nu\alpha\beta}p^{\alpha}
\mathcal{A}^{(1)\beta} +\frac{1}{2m}\,\partial_{x[\mu}\mathcal{V}_{\nu]}^{(0)}
-\frac{1}{m}\mathrm{Re\, Tr}\left(\sigma_{\mu\nu}I_{\mathrm{coll}}^{(1)}\right)\;.
\label{eq:sol-a1}
\end{eqnarray}
Note that the terms $\mathcal{A}_{\mu}^{(0)}$, $\mathcal{S}_{\nu\mu}^{(0)}$,
$\mathcal{V}_{\nu}^{(0)}$, and $\mathrm{Tr}\left(\Gamma_{a}I_{\mathrm{coll}}^{(1)}\right)$
are all on-shell. We can act with $(p^{2}-m^{2})$ onto the above equations and
verify that the resulting equations are consistent with Eqs.\ (\ref{eq:mass-shell_f1}) --
(\ref{eq:mass-shell_v1}) for the on-shell conditions.

From Eqs.\ (\ref{eq:mass-shell_f1}), (\ref{eq:mass-shell_a1}), (\ref{eq:mass-shell_s1}),
and (\ref{eq:mass-shell_v1}) we observe that $\mathcal{F}^{(1)}$,
$\mathcal{V}_{\mu}^{(1)}$, $\mathcal{A}^{(1)}_{\mu}$, and $\mathcal{S}_{\mu\nu}^{(1)}$
in principle have off-shell contributions at
$\mathcal{O}(\hbar)$, while $\mathcal{P}^{(1)}$ remains on-shell at
$\mathcal{O}(\hbar)$ as shown in Eq.\ (\ref{eq:mass-shell_p1}), i.e.,
off-shell effects should enter at most at $\mathcal{O}(\hbar^{2})$.

\subsection{Second order in $\hbar$}

At order $\mathcal{O}(\hbar^{2})$ the collision term (\ref{eq:collision_term}) reads
\begin{equation}
I_{\mathrm{coll}}^{(2)}  \equiv  \Delta I_{\mathrm{coll}}^{(1)}
+ I_{\mathrm{coll, PB}}^{(0)} \; ,
\label{eq:def_I2}
\end{equation}
where
\begin{equation}
\Delta I_{\mathrm{coll}}^{(1)}  \equiv  -\frac{i}{2} \left[ \Sigma^{<(1)}(x,p)G^{>(0)}(x,p)
-\Sigma^{>(1)}(x,p)G^{<(0)}(x,p) +\Sigma^{<(0)}(x,p)G^{>(1)}(x,p)
-\Sigma^{>(0)}(x,p)G^{<(1)}(x,p) \right] \label{eq:def_DeltaI1}
\end{equation}
arises from an expansion to first order in $\hbar$ of the first term in
Eq.\ (\ref{eq:collision_term}), while
\begin{equation}
I_{\mathrm{coll, PB}}^{(0)}  \equiv  -\frac{1}{4} \left[
\left\{ \Sigma^{<(0)}(x,p),G^{>(0)}(x,p)\right\} _{\mathrm{PB}}
-\left\{ \Sigma^{>(0)}(x,p),G^{<(0)}(x,p)\right\}_{\mathrm{PB}}\right]\;,\label{eq:def_IPB}
\end{equation}
is the leading-order contribution from the second term in Eq.\ (\ref{eq:collision_term}).
Note that all two-point functions appearing in $\Sigma^{\gtrless(1)}(x,p)$ are
taken at zeroth order in $\hbar$, $G^{\gtrless} \equiv G^{\gtrless(0)} $, except for one,
for which one has to take the first-order contribution $G^{\gtrless(1)}$.


Equations (\ref{eq:scalar_real}) -- (\ref{eq:tensor_real}) read to second order in $\hbar$:
\begin{eqnarray}
p^{\mu}\mathcal{V}_{\mu}^{(2)}-m\mathcal{F}^{(2)}
& = & \mathrm{Re\, Tr}\left(I_{\mathrm{coll}}^{(2)}\right)\;,\label{eq:scalar_real_2} \\
\frac{1}{2}\, \partial^{\mu}_{x}\mathcal{A}^{(1)}_{\mu}+m\mathcal{P}^{(2)}
& = & \mathrm{Re\, Tr}\left(i\gamma^{5}I_{\mathrm{coll}}^{(2)}\right)\;,
\label{eq:pseudoscalar_real_2} \\
\frac{1}{2}\, \partial_{x}^{\nu}\mathcal{S}_{\nu\mu}^{(1)}- p_{\mu}\mathcal{F}^{(2)}
+ m\mathcal{V}_{\mu}^{(2)}
& = & - \mathrm{Re\, Tr}\left(\gamma_{\mu}I_{\mathrm{coll}}^{(2)}\right)\;,
\label{eq:vector_real_2} \\
- \frac{1}{2}\, \partial_{x,\mu} \mathcal{P}^{(1)}
+ \frac{1}{2}\, \epsilon_{\mu\nu\alpha\beta}p^{\nu}
\mathcal{S}^{(2)\alpha\beta} +m\mathcal{A}_{\mu}^{(2)}
& = & \mathrm{Re\,Tr}\left(\gamma^{5}\gamma_{\mu}I_{\mathrm{coll}}^{(2)}\right)\;,
\label{eq:axialvector_real_2} \\
\frac{1}{2}\,\partial_{x[\mu}\mathcal{V}_{\nu]}^{(1)}
-\epsilon_{\mu\nu\alpha\beta}p^{\alpha}\mathcal{A}^{(2)\beta}
-m\mathcal{S}_{\mu\nu}^{(2)}
& = & \mathrm{Re\, Tr}\left(\sigma_{\mu\nu}I_{\mathrm{coll}}^{(2)}\right)\;.
\label{eq:tensor_real_2}
\end{eqnarray}
On the other hand, Eqs.\ (\ref{eq:scalar_im}) -- (\ref{eq:tensor_im}) read to second
order in $\hbar$:
\begin{eqnarray}
\frac{1}{2}\, \partial^{\mu}_{x}\mathcal{V}^{(1)}_{\mu}
& = & \mathrm{Im\, Tr}\left(I_{\mathrm{coll}}^{(2)}\right)\;,\label{eq:scalar_im_2} \\
p^\mu \mathcal{A}^{(2)}_\mu
& = & \mathrm{Im\, Tr}\left(-i \gamma^{5}I_{\mathrm{coll}}^{(2)}\right)\;,
\label{eq:pseudoscalar_im_2} \\
\frac{1}{2}\,\partial_{x,\mu}\mathcal{F}^{(1)}+ p^{\nu}\mathcal{S}_{\nu\mu}^{(2)}
& = & \mathrm{Im\, Tr}\left(\gamma_{\mu}I_{\mathrm{coll}}^{(2)}\right)\;,
\label{eq:vector_im_2} \\
\frac{1}{4}\, \epsilon_{\mu\nu\alpha\beta} \partial_{x}^{\nu}\mathcal{S}^{(1)\alpha\beta}
+ p_{\mu}\mathcal{P}^{(2)}
& = & \mathrm{Im \, Tr}\left(\gamma^{5}\gamma_{\mu}I_{\mathrm{coll}}^{(2)}\right)\;,
\label{eq:axialvector_im_2} \\
\frac{1}{2}\,\epsilon_{\mu\nu\alpha\beta}\partial_{x}^{\alpha}
\mathcal{A}^{(1)\beta} + p_{[\mu}\mathcal{V}_{\nu]}^{(2)}
& = & -\mathrm{Im\,Tr}\left(\sigma_{\mu\nu}I_{\mathrm{coll}}^{(2)}\right)\;.
\label{eq:tensor_im_2}
\end{eqnarray}

Taking the four-divergence of Eq.\ (\ref{eq:vector_real_1}) and using
Eq.\ (\ref{eq:scalar_im_2})
we derive a Boltzmann equation for the first-order scalar component at
$\mathcal{O}(\hbar^2)$,
\begin{equation}
p\cdot\partial_{x}\mathcal{F}^{(1)}  = 2m\, \mathrm{Im\,Tr}\left(I_{\mathrm{coll}}^{(2)}\right)
+ \mathrm{Re\,Tr}\left(\gamma \cdot \partial_{x}I_{\mathrm{coll}}^{(1)}\right)\;.
\label{eq:boltzmann_scalar-2nd_v1}
\end{equation}
Alternatively, we can obtain a Boltzmann equation for $\mathcal{F}^{(1)}$ by
multiplying Eq.\ (\ref{eq:vector_im_2}) with $p^\mu$,
\begin{equation}\label{eq:boltzmann_scalar-2nd_v2}
p\cdot\partial_{x}\mathcal{F}^{(1)}
= 2p^\mu \, \mathrm{Im\,Tr}\left(\gamma_\mu I_{\mathrm{coll}}^{(2)}\right) \;.
\end{equation}
Comparing the right-hand sides of Eqs.\ (\ref{eq:boltzmann_scalar-2nd_v1}) and
(\ref{eq:boltzmann_scalar-2nd_v2}), we derive a constraint equation for the
collision terms at first and second order,
\begin{equation}
\mathrm{Re\, Tr}\left(\gamma\cdot \partial_{x}I_{\mathrm{coll}}^{(1)}\right)  =
2 \, \mathrm{Im\,Tr}\left[(\gamma\cdot p-m)I_{\mathrm{coll}}^{(2)}\right]\;.
\label{eq:constraint-i2_1}
\end{equation}

Multiplying Eq.\ (\ref{eq:tensor_im_2}) with $p_\sigma \epsilon^{\rho \sigma \mu \nu}$ and
using Eq.\ (\ref{eq:pseudoscalar_im_1}) we derive a Boltzmann equation for the first-order
axial-vector component at $\mathcal{O}(\hbar^{2})$,
\begin{eqnarray}
p\cdot\partial_{x}\mathcal{A}^{(1)\mu} &=& -\epsilon^{\mu\nu\alpha \beta} p_{\nu}\,
\mathrm{Im\,Tr}\left(\sigma_{\alpha\beta}I_{\mathrm{coll}}^{(2)}\right)
-\mathrm{Re\, Tr}\left(\gamma^{5}\partial_{x}^{\mu}
I_{\mathrm{coll}}^{(1)}\right)\nonumber\\
& = &  - 2p^\mu \,\mathrm{Im\, Tr}\left( \gamma^5 I_{\mathrm{coll}}^{(2)} \right)
- 2\,  \mathrm{Im \, Tr} \left( \gamma \cdot p\, \gamma^5 \gamma^\mu
I_{\mathrm{coll}}^{(2)} \right)
-\mathrm{Re\, Tr}\left(\gamma^{5}\partial_{x}^{\mu} I_{\mathrm{coll}}^{(1)}\right)\;.
\label{eq:boltzmann_axialvector-2nd_v1}
\end{eqnarray}
An alternative way of deriving a Boltzmann equation for $\mathcal{A}^{(1)\mu}$ is
by multiplying Eq.\ (\ref{eq:tensor_real_1}) with
$\epsilon^{\rho\sigma \mu \nu} \partial_{x,\sigma}$ and using
Eqs.\ (\ref{eq:pseudoscalar_real_2}) and (\ref{eq:axialvector_im_2}),
\begin{equation}\label{eq:boltzmann_axialvector-2nd_v2}
p\cdot\partial_{x}\mathcal{A}^{(1)\mu}
= - 2 p^\mu\, \mathrm{Im \, Tr} \left( \gamma^5 I_{\mathrm{coll}}^{(2)}
\right) - 2m\,  \mathrm{Im \, Tr} \left( \gamma^5 \gamma^\mu I_{\mathrm{coll}}^{(2)} \right)
- \frac 12 \epsilon^{\mu \nu \alpha \beta} \mathrm{Re\, Tr}
\left(\sigma_{\alpha \beta} \partial_{x,\nu} I_{\mathrm{coll}}^{(1)}\right)\;.
\end{equation}
Comparing the right-hand sides of Eqs.\ (\ref{eq:boltzmann_axialvector-2nd_v1}) and
(\ref{eq:boltzmann_axialvector-2nd_v2}) we derive another constraint equation for the
collision terms at first and second order,
\begin{equation}
\frac 12\epsilon^{\mu\nu \alpha \beta} \, \mathrm{Re\, Tr}
\left(\sigma_{\alpha \beta} \partial_{x,\nu} I_{\mathrm{coll}}^{(1)}\right)
- \mathrm{Re\, Tr}\left(\gamma^5 \partial_{x}^\mu I_{\mathrm{coll}}^{(1)}\right)  =
2 \, \mathrm{Im\,Tr}\left[(\gamma\cdot p-m) \gamma^5 \gamma^\mu
I_{\mathrm{coll}}^{(2)}\right]\;.
\label{eq:constraint-i2_2}
\end{equation}
With the help of Eq.\ (\ref{another-eq-coll}), one can prove that
Eqs.\ (\ref{eq:constraint-i2_1}), (\ref{eq:constraint-i2_2})
are satisfied up to terms of order $\mathcal{O}(\hbar G_c^4)$.

One can also derive Boltzmann-type equations similar to
Eqs.\ (\ref{eq:boltzmann_scalar-2nd_v1}),
(\ref{eq:boltzmann_axialvector-2nd_v1}), or Eqs.\ (\ref{eq:boltzmann_scalar-2nd_v2}),
(\ref{eq:boltzmann_axialvector-2nd_v2}), respectively, for the other components of
the Wigner functions, as well as additional constraints similar to
Eqs.\ (\ref{eq:constraint-i2_1}), (\ref{eq:constraint-i2_2}).
However, we aim at deriving a Boltzmann-type equation
for the matrix-valued spin distribution function, and for this purpose
we will only make use of
Eqs.\ (\ref{eq:boltzmann_scalar-2nd_v1}), (\ref{eq:boltzmann_axialvector-2nd_v1}).

\section{Matrix-valued spin distribution functions and quasi-classical
contributions to Wigner functions}

\label{sec:spin-dist}

To the order we are computing, the Wigner functions
$G^{\gtrless(0)}$ and $G^{\gtrless(1)}$ enter the collision terms
$I_{\mathrm{coll}}^{(1)}$, $I_{\mathrm{coll}}^{(2)}$ in the Boltzmann
equations (\ref{eq:boltzmann_scalar-1st}), (\ref{eq:boltzmann_axialvector-1st}),
(\ref{eq:boltzmann_scalar-2nd_v1}), (\ref{eq:boltzmann_scalar-2nd_v2}),
(\ref{eq:boltzmann_axialvector-2nd_v1}), and (\ref{eq:boltzmann_axialvector-2nd_v2}),
cf.\ Eqs.\ (\ref{eq:collision-1st}), (\ref{eq:def_DeltaI1}), and
(\ref{eq:def_IPB}). The zeroth-order quantities $G^{\gtrless(0)}$
and their components are all on-shell (see Sec.\ \ref{zeroth-order-h-expand}
for $G^{<(0)}$; $G^{>(0)}$ has the same properties) and take the
same form as in kinetic theory in the absence of
collisions \cite{Weickgenannt:2019dks}.
In the following, we call a contribution to
the $n$-th order Wigner function $G^{\gtrless(n)}$ a \emph{quasi-classical} (qc)
contribution, if its components satisfy
Eqs.\ (\ref{eq:pseudoscalar_real_0a}) -- (\ref{eq:tensor_real_0a}).
However, the Clifford components of the first-order quantities $G^{\gtrless(1)}$ satisfy
Eqs.\ (\ref{eq:sol-p1}) -- (\ref{eq:sol-a1}),
which contain additional terms besides the
(first-order) quasi-classical contributions. These are \emph{gradients} of zeroth-order
quantities as well as \emph{collision terms}. We denote these
additional contributions with an index $\nabla$.
In general, the Wigner function at $\mathcal{O}(\hbar)$ can be written
as follows:
\begin{equation}
G^{\lessgtr(1)}=G_{\mathrm{qc}}^{\lessgtr(1)}
+G_{\nabla}^{\lessgtr(1)}+G_{\mathrm{off}}^{\lessgtr(1)}\;,
\label{eq:wig-kin-nonkin-off}
\end{equation}
where $G_{\mathrm{qc}}^{\lessgtr(1)}$, $G_{\nabla}^{\lessgtr(1)}$,
and $G_{\mathrm{off}}^{\lessgtr(1)}$ denote the quasi-classical term,
the on-shell gradient and collision term,
and the off-shell term, respectively. Note that $G_{\mathrm{qc}}^{\lessgtr(1)}$ and
$G_{\nabla}^{\lessgtr(1)}$ are both on-shell. By definition,
the Clifford components of the quasi-classical contribution $G_{\mathrm{qc}}^{<(1)}$
satisfy
\begin{eqnarray}
\mathcal{P}_{\mathrm{qc}}^{(1)} & = & 0\;,\label{eq:1st-kin-comp_P} \\
\mathcal{V}_{{\mathrm{qc}},\mu}^{(1)}
& = & \frac{1}{m}p_{\mu}\mathcal{F}^{(1)}_{\mathrm{qc}}\;,\label{eq:1st-kin-comp_V} \\
\mathcal{S}_{{\mathrm{qc}},\mu\nu}^{(1)}
& = & -\frac{1}{m}\epsilon_{\mu\nu\alpha\beta}p^{\alpha}
\mathcal{A}^{(1)\beta}_{\mathrm{qc}}\;,
\label{eq:1st-kin-comp_S}
\end{eqnarray}
where $\mathcal{F}^{(1)}_{\mathrm{qc}}$ and $\mathcal{A}^{(1)\beta}_{\mathrm{qc}}$
are the quasi-classical contributions of the respective Clifford components
(which are on-shell). On the other hand,
the gradient and collision contributions contain
space-time derivatives of the zeroth-order Wigner functions
and the collision term $I_{\mathrm{coll}}^{(1)}$,
\begin{eqnarray}
\mathcal{P}_\nabla^{(1)}
& = & -\frac{1}{2m}\partial_{x}^{\mu}\mathcal{A}_{\mu}^{(0)}
+\frac{1}{m}\mathrm{Re\,Tr}\left(i\gamma^{5}I_{\mathrm{coll}}^{(1)}\right)\;,
\label{eq:1st-nonkin-comp_P} \\
\mathcal{V}_{\nabla,\mu}^{(1)}
& = & -\frac{1}{2m}\partial_{x}^{\nu}\mathcal{S}_{\nu\mu}^{(0)}
-\frac{1}{m}\mathrm{Re\,Tr}\left(\gamma_{\mu}I_{\mathrm{coll}}^{(1)}\right)\;,
\label{eq:1st-nonkin-comp_V} \\
\mathcal{S}_{\nabla,\mu\nu}^{(1)}
& = & \frac{1}{2m}\,\partial_{x[\mu}\mathcal{V}_{\nu]}^{(0)}
-\frac{1}{m}\mathrm{Re\,Tr}\left(\sigma_{\mu\nu}I_{\mathrm{coll}}^{(1)}\right)\;.
\label{eq:1st-nonkin-comp_S}
\end{eqnarray}
Note that we have set $\mathcal{F}_\nabla^{(1)}=0$ and
$\mathcal{A}_\nabla^{(1)\mu}=0$.
If $\mathcal{F}_\nabla^{(1)}$ and $\mathcal{A}_\nabla^{(1)\mu}$ are nonvanishing, it
can be proved that they can be absorbed into the quasi-classical contributions of
the Wigner functions
by a redefinition of the first-order matrix-valued spin distribution functions
(see discussion below).
Finally, the \emph{off-shell contributions} are defined as
\begin{eqnarray}
\mathcal{P}_\mathrm{off}^{(1)}
& = & 0\;, \label{eq:1st-off-comp_P} \\
\mathcal{V}_{\mathrm{off},\mu}^{(1)}
& = & \frac{1}{m} p_\mu\, \mathcal{F}^{(1)}_\mathrm{off} \;,
\label{eq:1st-off-comp_V} \\
\mathcal{S}_{\mathrm{off},\mu\nu}^{(1)}
& = & -\frac{1}{m}\epsilon_{\mu\nu\alpha\beta}p^{\alpha}
\mathcal{A}^{(1)\beta}_\mathrm{off}\;.
\label{eq:1st-off-comp_S}
\end{eqnarray}
The sum of quasi-classical, gradient and collision, and off-shell contributions
(\ref{eq:1st-kin-comp_P}) -- (\ref{eq:1st-off-comp_S}) satisfy
Eqs.\ (\ref{eq:sol-p1}) -- (\ref{eq:sol-a1}).
Since $G^{<(0)}$ contains only quasi-classical contributions, there is no need to
specify zeroth-order quantities with the subscript ``qc''.

In the remainder of this section we will analyze the
structure of the quasi-classical as well as gradient and collision
contributions in terms of single-particle distribution
functions, which are $(2\times2)$ matrices in the space of spin indices, the so-called
matrix-valued single-particle distribution functions (MVSDs).
Since these contributions exist also in the absence of collisions,
the starting point is the solution of the free Dirac equation, which
can be expanded in the standard way as
\begin{eqnarray}
\psi(x) & = & \sum_{s}\int\frac{d^{3}\mathbf{k}}{(2\pi\hbar)^{3}}\,\frac{1}{2E_{k}}
\left[a(s,\mathbf{k})u(s,\mathbf{k})e^{-ik\cdot x/\hbar}
+b^{\dagger}(s,\mathbf{k})v(s,\mathbf{k})e^{ik\cdot x/\hbar}\right]\;,
\label{eq:field-expansion_psi}\\
\overline{\psi}(x) & = & \sum_{s}\int\frac{d^{3}\mathbf{k}}{(2\pi\hbar)^{3}}\,\frac{1}{2E_{k}}
\left[a^{\dagger}(s,\mathbf{k})\overline{u}(s,\mathbf{k})e^{ik\cdot x/\hbar}
+b(s,\mathbf{k})\overline{v}(s,\mathbf{k})e^{-ik\cdot x/\hbar}\right]\;,
\label{eq:field-expansion_barpsi}
\end{eqnarray}
where $k^{0}\equiv E_{k}\equiv E_{\mathbf{k}}=\sqrt{\mathbf{k}^{2}+m^{2}}$,
the index $s=\pm1$ denotes the spin state parallel or anti-parallel
to the quantization direction $\mathbf{n}$, and $a(s,\mathbf{k})$,
$a^{\dagger}(s,\mathbf{k})$, $b(s,\mathbf{k})$, and
$b^{\dagger}(s,\textbf{\ensuremath{\mathbf{k}}})$
are creation and annihilation operators for fermions
and antifermions (of dimension {[}energy $\times$ length$^{3}${]}$^{1/2}$),
respectively. The spinors for fermions and antifermions are
$u(\mathbf{k},s)$ and $v(\mathbf{k},s)$, respectively. In Dirac representation,
the spinors take the explicit form
\begin{eqnarray}
u(s,\mathbf{p}) & = & \sqrt{E_{p}+m}\left(\begin{array}{c}
\chi_{s}\\
\frac{\boldsymbol{\sigma}\cdot\mathbf{p}}{E_{p}+m}\chi_{s}
\end{array}\right)\;,\nonumber \\
v(s,\mathbf{p}) & = & -i\sqrt{E_{p}+m}\left(\begin{array}{c}
\frac{\boldsymbol{\sigma}\cdot\mathbf{p}}{E_{p}+m}\sigma_{2}\chi_{s}^{*}\\
\sigma_{2}\chi_{s}^{*}
\end{array}\right)\;.\label{eq:u-v-2nd-quant}
\end{eqnarray}
They satisfy the relation $v(s,\mathbf{p})=i\gamma^{2}u^{*}(s,\mathbf{p})$,
which one readily proves using
$\sigma_{2}\boldsymbol{\sigma}\sigma_{2}=-\boldsymbol{\sigma}^{*}$.
Here, $\chi_{s}$ is the Pauli spinor with spin projection $s$ quantized
along the direction $\mathbf{n}=(\sin\theta\cos\phi,\sin\theta\sin\phi,\cos\theta)^{T}$,
so we have $(\mathbf{n}\cdot\boldsymbol{\sigma})\chi_{s}=s\chi_{s}$.
One can verify that the spin orientation of the state $\sigma_{2}\chi_{s}^{*}$
is opposite to that of $\chi_{s}$.

We now insert Eqs.\ (\ref{eq:field-expansion_psi}) and (\ref{eq:field-expansion_barpsi})
into Eq.\ (\ref{eq:wigner-def}) and, neglecting contributions from space-like
momenta $p^{2}<0$ \cite{DeGroot:1980dk}, obtain the Wigner function,
\begin{eqnarray}
G_{\alpha\beta}^{<}(x,p) & = & -(2\pi\hbar)^{2}\sum_{r,s}\int\frac{d^{4}q}{(2\pi\hbar)^{4}}\,
\frac{1}{4E_{\mathbf{p}+\mathbf{q}/2}E_{\mathbf{p}-\mathbf{q}/2}}\nonumber \\
 &  & \times\left[\delta\left(p_{0}-\frac{E_{\mathbf{p}+\mathbf{q}/2}
 +E_{\mathbf{p}-\mathbf{q}/2}}{2}\right)\,\delta\left(q^{0}-E_{\mathbf{p}
 +\mathbf{q}/2}+E_{\mathbf{p}-\mathbf{q}/2}\right)\,e^{-iq\cdot x/\hbar}\right.\nonumber \\
 &  & \times\;u_{\alpha}\left(r,\mathbf{p}+\frac{\mathbf{q}}{2}\right)\,
 \overline{u}_{\beta}\left(s,\mathbf{p}-\frac{\mathbf{q}}{2}\right)
 \left\langle a^{\dagger}\left(s,\mathbf{p}-\frac{\mathbf{q}}{2}\right)\,
 a\left(r,\mathbf{p}+\frac{\mathbf{q}}{2}\right)\right\rangle \nonumber \\
 &  & +\delta\left(p_{0}+\frac{E_{\mathbf{p}+\mathbf{q}/2}+E_{\mathbf{p}
 -\mathbf{q}/2}}{2}\right)\,\delta\left(q^{0}+E_{\mathbf{p}+\mathbf{q}/2}
 -E_{\mathbf{p}-\mathbf{q}/2}\right)\,e^{iq\cdot x/\hbar}\nonumber \\
 &  & \times\left.v_{\alpha}\left(s,-\mathbf{p}+\frac{\mathbf{q}}{2}\right)\,
 \overline{v}_{\beta}\left(r,-\mathbf{p}-\frac{\mathbf{q}}{2}\right)
 \left\langle b\left(r,-\mathbf{p}-\frac{\mathbf{q}}{2}\right)
 b^{\dagger}\left(s,-\mathbf{p}+\frac{\mathbf{q}}{2}\right)\right\rangle \right]\;.
 \label{eq:Gsmaller_v1}
\end{eqnarray}
We see that the Wigner function depends on
$\left\langle a^{\dagger}\left(s,\mathbf{p}-\mathbf{q}/2\right)\,
a\left(r,\mathbf{p}+\mathbf{q}/2\right)\right\rangle $
and $\left\langle b\left(r,-\mathbf{p}-\mathbf{q}/2\right)
b^{\dagger}\left(s,-\mathbf{p}+\mathbf{q}/2\right)\right\rangle $,
a kind of density-matrix elements, which are in general nondiagonal
in spin and momentum space. The delta functions in Eq.\ (\ref{eq:Gsmaller_v1})
show that $G_{\alpha\beta}^{<}(x,p)$ has both on-shell and off-shell
contributions. While strictly valid only in the noninteracting case (with
the collision term $I_{\mathrm{coll}}$ set to zero), the result (\ref{eq:Gsmaller_v1})
still contains all orders of $\hbar$. For the collision term, we
require $G^{<}(x,p)$ up to first order in $\hbar$, i.e., we are
allowed to expand Eq.\ (\ref{eq:Gsmaller_v1}) up to this order.

In order to exhibit the structure of the on-shell part of
$G_{\alpha\beta}^{<}(x,p)$ in Eq.\ (\ref{eq:Gsmaller_v1}),
we expand the integrand in powers of $\mathbf{q}$,
with the exception of the creation and annihilation operators. Such an expansion
is equivalent to an expansion in $\hbar$,
because $\mathbf{q}$, accompanied
by a factor $e^{i \mathbf{q} \cdot \mathbf{x}/\hbar}$, is
equivalent to $-i\hbar \nabla_x$ acting on the latter exponential.
At leading order $\mathcal{O}(\hbar^{0})$,
it can be shown that the Wigner function can be written as
\begin{eqnarray}
G_{\alpha\beta}^{<(0)}(x,p) & = & -2\pi\hbar\,\theta(p_{0})\delta\left(p^{2}-m^{2}\right)
\sum_{r,s}u_{\alpha}\left(r,p\right)\overline{u}_{\beta}\left(s,p\right)\,
f_{sr}^{(+,0)}\left(x,p\right)\nonumber \\
 &  & -2\pi\hbar\,\theta(-p_{0})\delta\left(p^{2}-m^{2}\right)\sum_{r,s}
 v_{\alpha}\left(s,\overline{p}\right)\overline{v}_{\beta}\left(r,\overline{p}\right)
 \left[\delta_{sr}-f_{sr}^{(-,0)}\left(x,\overline{p}\right)\right]\;,\label{eq:g-less-0}
\end{eqnarray}
where the MVSDs for fermions and antifermions are defined as
\begin{eqnarray}
f_{sr}^{(+,0)}\left(x,p\right) & \equiv & \int\frac{d^{4}q}{2(2\pi\hbar)^{4}}\,2\pi\hbar\,
\delta(p\cdot q)\,e^{-iq\cdot x/\hbar}\,
\left\langle a^{\dagger}\left(s,\mathbf{p}-\frac{\mathbf{q}}{2}\right)\,
a\left(r,\mathbf{p}+\frac{\mathbf{q}}{2}\right)\right\rangle \;,\label{eq:f_rs-distr_+}\\
f_{sr}^{(-,0)}\left(x,\overline{p}\right) & \equiv & \int\frac{d^{4}q}{2(2\pi\hbar)^{4}}\,
2\pi\hbar\,\delta(\overline{p}\cdot q)\,e^{-iq\cdot x/\hbar}\,
\left\langle b^{\dagger}\left(s,-\mathbf{p}-\frac{\mathbf{q}}{2}\right)
b\left(r,-\mathbf{p}+\frac{\mathbf{q}}{2}\right)\right\rangle \;,\label{eq:f_rs-distr_-}
\end{eqnarray}
with $p\equiv(E_{\mathbf{p}},\mathbf{p})$ and
$\overline{p}^{\mu}\equiv(E_{\mathbf{p}},-\mathbf{p})$.
Note that the overall factor $\hbar$ in Eq.\ (\ref{eq:g-less-0}) is convention and
does not participate in the power counting in $\hbar$ (it is the same factor that
is factored out on both sides of the KB equation, see discussion in Sec.\ \ref{sec:kbA}).
One can verify that $G^{<(0)}$ in Eq.\ (\ref{eq:g-less-0}) takes the
same form as in kinetic theory without collisions. Note that even
in the noninteracting case without collisions, there are contributions
of $\mathcal{O}(\hbar)$ to the Wigner function (\ref{eq:Gsmaller_v1}) in the
$\mathbf{q}$-expansion, which result in space-time derivatives of
$f_{sr}^{(\pm,0)}$. With collisions, the Wigner function of $\mathcal{O}(\hbar)$
has to be determined by solving the respective
equation of motion at the same order.

We assume that the quasi-classical contribution to
the Wigner function at $\mathcal{O}(\hbar)$ arises from the
$\mathcal{O}(\hbar)$ corrections of the MVSDs to Eq.\ (\ref{eq:g-less-0}),
\begin{eqnarray}
G_{\mathrm{qc},\alpha\beta}^{<(1)}(x,p) & = &
-2\pi\hbar\,\theta(p_{0})\delta\left(p^{2}-m^{2}\right)
\sum_{r,s}u_{\alpha}\left(r,p\right)\overline{u}_{\beta}\left(s,p\right)\,
f_{sr}^{(+,1)}\left(x,p\right)\nonumber \\
 &  & +2\pi\hbar\,\theta(-p_{0})\delta\left(p^{2}-m^{2}\right)\sum_{r,s}
 v_{\alpha}\left(s,\overline{p}\right)\overline{v}_{\beta}\left(r,\overline{p}\right)
 f_{sr}^{(-,1)}\left(x,\overline{p}\right)\;.\label{eq:g-less-1}
\end{eqnarray}
Note that $f_{sr}^{(\pm,0)}$ and $f_{sr}^{(\pm,1)}$ are to be determined
by solving the Boltzmann equation at $\mathcal{O}(\hbar)$ and $\mathcal{O}(\hbar^{2})$,
respectively. One can verify that the Clifford components of $G^{<(1)}_\mathrm{qc}$
in Eq.\ (\ref{eq:g-less-1}) satisfy Eqs.\ (\ref{eq:1st-kin-comp_P})
-- (\ref{eq:1st-kin-comp_S}).
Therefore, up to $\mathcal{O}(\hbar)$ the quasi-classical contribution to the
Wigner function can be written as
\begin{eqnarray}
G_{\mathrm{qc}, \alpha\beta}^{<}(x,p) & = & G_{\alpha\beta}^{<(0)}(x,p)
+\hbar G_{\mathrm{qc}, \alpha\beta}^{<(1)}(x,p)+\mathcal{O}(\hbar^{2})\nonumber \\
 & = & -2\pi\hbar\,\theta(p_{0})\delta\left(p^{2}-m^{2}\right)\sum_{r,s}
 u_{\alpha}\left(r,p\right)\overline{u}_{\beta}\left(s,p\right)\,f_{sr}^{(+)}\left(x,p\right)
 \nonumber \\
 &  & -2\pi\hbar\,\theta(-p_{0})\delta\left(p^{2}-m^{2}\right)\sum_{r,s}
 v_{\alpha}\left(s,\overline{p}\right)\overline{v}_{\beta}\left(r,\overline{p}\right)
 \left[\delta_{sr}-f_{sr}^{(-)}\left(x,\overline{p}\right)\right]+\mathcal{O}(\hbar^{2})\;,
 \label{eq:g-less-all}
\end{eqnarray}
where
\begin{equation}
f_{sr}^{(\pm)}\left(x,p\right)=f_{sr}^{(\pm,0)}\left(x,p\right)
+\hbar f_{sr}^{(\pm,1)}\left(x,p\right)
+\mathcal{O}(\hbar^{2})\;.\label{eq:mvsd-expand}
\end{equation}
Equation (\ref{eq:g-less-all}) indicates that, up to $\mathcal{O}(\hbar)$,
the quasi-classical part of the Wigner function can be expressed in terms of
the MVSDs.

On the other hand, the Wigner function $G^{>(0)}(x,p)$ can be
obtained from $G^{<(0)}(x,p)$ by replacing
$f_{sr}^{(+,0)}\rightarrow-(\delta_{sr}-f_{sr}^{(+,0)})$
and $\delta_{sr}-f_{sr}^{(-,0)}\rightarrow-f_{sr}^{(-,0)}$ in Eq.\
(\ref{eq:g-less-0}). At $\mathcal{O}(\hbar)$ one can convince oneself by an
explicit calculation that $G_{\mathrm{qc}}^{>(1)}(x,p)
=G_{\mathrm{qc}}^{<(1)}(x,p)$, i.e.,
it is also given by Eq.\ (\ref{eq:g-less-1}).

We can extract the MVSDs
$f_{sr}^{(\pm)}$ up to $\mathcal{O}(\hbar)$ by contracting
$G^{<}_\mathrm{qc}(x,p)$ in Eq.\ (\ref{eq:g-less-all}) with Dirac spinors and
an integration over $p^{0}$:
\begin{eqnarray}
f_{sr}^{(+)}(x,p) & = & -\frac{E_{p}}{4\pi\hbar m^{2}}\int dp^{0}\,\theta(p^{0})\,
\overline{u}(r,p)G_\mathrm{qc}^{<}(x,p)u(s,p)+\mathcal{O}(\hbar^{2})\;,\label{eq:g_sr+}\\
f_{sr}^{(-)}(x,\overline{p}) & = & \frac{E_{p}}{4\pi\hbar m^{2}}\int dp^{0}\,\theta(-p^{0})\,
\overline{v}(s,\overline{p})G_\mathrm{qc}^{<}(x,p)v(r,\overline{p})
+\delta_{rs}+\mathcal{O}(\hbar^{2})\;.
\label{eq:g_sr-}
\end{eqnarray}

We can also relate the MVSDs to the components
of the Clifford decomposition of $G^{<}(x,p)$. From
$\mathcal{F}(x,p)=\mathrm{Tr}[G^{<}(x,p)]$
and Eqs.\ (\ref{eq:g-less-0}), (\ref{eq:g-less-1}) we obtain the quasi-classical
part of the scalar component up to $\mathcal{O}(\hbar)$
\begin{eqnarray}
\mathcal{F}_\mathrm{qc} (x,p) & = & -2\pi\hbar\frac{m}{E_{p}}\left\{ \delta(p_{0}-E_{p})\,
\mathrm{tr}\left[f^{(+)}(x,p)\right]+\delta(p_{0}+E_{p})\,
\mathrm{tr}\left[f^{(-)}(x,\overline{p})-1\right]\right\} +\mathcal{O}(\hbar^{2})\;,
\label{eq:F_0}
\end{eqnarray}
where ``tr'' denotes the trace in the two-dimensional space of spin
indices and $f^{(+)}$ denotes the MVSD in $(2\times2)$ matrix form.
The $\hbar$ in the prefactor of the first term again does not participate
in the power counting in $\hbar$.
On the other hand, from $\mathcal{A}^{\mu}(x,p)
=\mathrm{Tr}[\gamma^{\mu}\gamma^{5}G^{<}(x,p)]$
and Eqs.\ (\ref{eq:g-less-0}), (\ref{eq:g-less-1}) we obtain the quasi-classical
part of the axial-vector component up to $\mathcal{O}(\hbar)$
\begin{eqnarray}
\mathcal{A}_\mathrm{qc}^{\mu} & = & -2\pi\hbar\frac{m}{E_{p}}\left\{ \delta(p_{0}-E_{p})
\,n_{j}^{(+)\mu}\,\mathrm{tr}\left[\tau_{j}^{T}\,f^{(+)}(x,p)\right]
+\delta(p_{0}+E_{p})\,n_{j}^{(-)\mu}\,\mathrm{tr}\left[\tau_{j}^{T}\,f^{(-)}(x,\overline{p})\right]
\right\} +\mathcal{O}(\hbar^{2})\;,\label{eq:A_0}
\end{eqnarray}
where $n_{j}^{(\pm)\mu}$, with $j=1,2,3$, are polarization four-vectors
defined by
\begin{equation}
n_{j}^{(\pm)\mu}\equiv n^{\mu}(\pm\mathbf{p},\mathbf{n}_{j})
=\left(\pm\frac{\mathbf{n}_{j}\cdot\mathbf{p}}{m},\mathbf{n}_{j}
+\frac{(\mathbf{n}_{j}\cdot\mathbf{p})\mathbf{p}}{m(E_{p}+m)}\right)^{T}\;.
\label{eq:spin-direction-lab}
\end{equation}
Here, $\mathbf{n}_{j}$ ($j=1,2,3$) are the set of orthonormal vectors
(which form a right-handed basis in three spatial dimensions)
\begin{equation}
\mathbf{n}_{1}\equiv\left(\begin{array}{c}
\cos\phi\cos\theta\\
\sin\phi\cos\theta\\
-\sin\theta
\end{array}\right)\;,\;\;\mathbf{n}_{2}\equiv\left(\begin{array}{c}
-\sin\phi\\
\cos\phi\\
0
\end{array}\right)\;,\;\;\mathbf{n}_{3}\equiv\left(\begin{array}{c}
\cos\phi\sin\theta\\
\sin\phi\sin\theta\\
\cos\theta
\end{array}\right)\;,\label{eq:n_j}
\end{equation}
where $\mathbf{n}_{3}\equiv\mathbf{n}$ equals the spin quantization
direction in the rest frame of the particle. These vectors are constant,
i.e., the angles $\phi$ and $\theta$ point into a fixed direction
in space. With the Pauli spinors
\begin{eqnarray}
\chi_{+} & = & \left(\begin{array}{c}
e^{-i\phi}\cos(\theta/2)\\
\sin(\theta/2)
\end{array}\right),\nonumber \\
\chi_{-} & = & \left(\begin{array}{c}
-e^{-i\phi}\sin(\theta/2)\\
\cos(\theta/2)
\end{array}\right),
\end{eqnarray}
we have the identity
\begin{equation}
\chi_{r}^{\dagger}\boldsymbol{\sigma}\chi_{s}=\left(\begin{array}{cc}
\mathbf{n}_{3} & \mathbf{n}_{1}-i\mathbf{n}_{2}\\
\mathbf{n}_{1}+i\mathbf{n}_{2} & -\mathbf{n}_{3}
\end{array}\right)_{rs}=\mathbf{n}_{j}(\tau_{j})_{rs}=\mathbf{n}_{j}(\tau_{j}^{T})_{sr}\;,
\label{eq:identity_spinors}
\end{equation}
where $r,s=\pm$ denotes the spin state and
$\boldsymbol{\tau}=(\tau_{1},\tau_{2},\tau_{3})$
is the vector of Pauli matrices in spin index space. One can verify
$n_{j}^{+}\cdot n_{j}^{+}=n_{j}^{-}\cdot n_{j}^{-}=-1$ (no summation
over $j$) with $n_{j}^{\pm\mu}$ given by Eq.\ (\ref{eq:spin-direction-lab}).
The order $\mathcal{O}(\hbar)$ quasi-classical contributions to the
other components $\mathcal{P}(x,p)$, $\mathcal{V}^{\mu}(x,p)$,
and $\mathcal{S}^{\mu\nu}(x,p)$ can be obtained
from Eqs.\ (\ref{eq:1st-kin-comp_P}) -- (\ref{eq:1st-kin-comp_S}).

To conclude this section, we summarize the main result:
We define the quasi-classical contribution to the Wigner function as
that whose Clifford components satisfy Eqs.\  (\ref{eq:1st-kin-comp_P}) --
(\ref{eq:1st-kin-comp_S}). The quasi-classical scalar and axial-vector components are
treated as independent variables. All components that satisfy Eqs.\
(\ref{eq:1st-kin-comp_P}) -- (\ref{eq:1st-kin-comp_S}) can be assembled into
$G^{\lessgtr}$ in terms of MVSDs in the same way as in the free-streaming or
collisionless case. This does not mean that the MVSDs have the same values as in the
collisionless case: they are the solutions of Boltzmann equations including collisions.
In other words, in terms of the MVSDs, the quasi-classical contribution
$G^{\lessgtr}_\mathrm{qc}$ takes the same form as in the collisionless case, but
the MVSDs assume different values. We remark that, as an alternative to the MVSDs,
we can also define a scalar spin-dependent distribution
by extending the phase space and introducing a continuous spin variable.
The scalar spin-dependent distribution is equivalent to the MVSDs and allows to
combine the Boltzmann equations for the scalar and axial-vector components
into a single equation, for details see Refs.\
\cite{Weickgenannt:2020aaf,Weickgenannt:2021cuo}.

\section{Boltzmann equations with collisions at leading order}

\label{sec:boltzmann-leading}
In this section we will derive Boltzmann equations in terms of
the leading-order MVSDs,
based on Eqs.\ (\ref{eq:boltzmann_scalar-1st})
and (\ref{eq:boltzmann_axialvector-1st}) for the scalar and axial-vector components,
respectively. For the sake of simplicity, we will neglect the contribution
of antiparticles and consequently suppress the
``+'' superscripts at the MVSDs. As shown in Eq.\ (\ref{eq:g-less-0}),
the Wigner function and its components at $\mathcal{O}(\hbar^{0})$ are on-shell
and take the same form as in kinetic theory. Also the collision term
$I_{\mathrm{coll}}^{(1)}$ appearing in Boltzmann equations, i.e.,
Eqs.\ (\ref{eq:boltzmann_scalar-1st}), (\ref{eq:boltzmann_axialvector-1st}),
involves the Wigner functions only
at $\mathcal{O}(\hbar^{0})$, as shown in Eq.\
(\ref{eq:collision-1st}). Another feature of the leading-order Boltzmann
equations is that the collision term $I_{\mathrm{coll}}^{(1)}$ is
local in space-time, namely, space-time derivatives are absent,
which means that the particles collide at a single point in space-time.

In order to isolate the fermion contribution,
we integrate Eq.\ (\ref{eq:boltzmann_scalar-1st}) over $p_{0}$ from $0$ to $\infty$
and obtain, using Eq.\ (\ref{eq:F_0}) to lowest order in $\hbar$, the Boltzmann
equation for the trace of the MVSD,
\begin{equation}
\frac{1}{E_{p}}\,p\cdot\partial_{x}\mathrm{tr}\left[f^{(0)}(x,p)\right]
=-\frac{1}{\pi\hbar}\int_{0}^{\infty}dp_{0}\,
\mathrm{Im}\mathrm{Tr}\left(I_{\mathrm{coll}}^{(1)}\right)\;.
\label{eq:boltzmann-fs-1st-order}
\end{equation}
Similarly, we integrate Eq.\
(\ref{eq:boltzmann_axialvector-1st}) over $p_{0}$  from $0$ to $\infty$
to obtain the Boltzmann equation for the polarization part of the MVSD
\begin{equation}
\frac{1}{E_{p}}\,p\cdot\partial_{x}\mathrm{tr}\left[n_{j}^{(+)\mu}\tau_{j}^{T}
f^{(0)}(x,p)\right]=\frac{1}{2\pi\hbar m}\epsilon^{\mu\nu\alpha\beta}
\int_{0}^{\infty}dp_{0}\, p_{\nu}\mathrm{Im}\mathrm{Tr}\left(\sigma_{\alpha\beta}
I_{\mathrm{coll}}^{(1)}\right)\;.\label{eq:boltzmann-fs-1st-order-1}
\end{equation}
In the above two equations $f^{(0)}$ denotes the matrix form of
the zeroth-order MVSD, which can be expanded
in a basis of unit matrix and Pauli matrices as
\begin{equation} \label{eq:decomp_f0}
f^{(0)}=\frac{1}{2}\mathrm{tr}\left(f^{(0)}\right) 
+\frac{1}{2}\boldsymbol{\tau}^{T}\cdot\mathrm{tr}\left(\boldsymbol{\tau}^{T}f^{(0)}\right)\;.
\end{equation}
We observe that the trace and the polarization part of
the MVSD are equivalent to the MVSD itself.
Equations (\ref{eq:boltzmann-fs-1st-order}) and (\ref{eq:boltzmann-fs-1st-order-1})
constitute the Boltzmann equations at leading order in $\hbar$.

\begin{figure}
\caption{\label{fig:feynman}Feynman diagrams for (a,c) $\Sigma^{>}(x,p)$ and
(b,d) $\Sigma^{<}(x,p)$. Solid lines represent fermion propagators,
wavy lines represent boson propagators, solid circles represent the
vertices $g_{a}$ with $a$ labeling coupling types. In the NJL model
wavy lines attached with two solid circles represent vertices $2G_{a}$.}
\includegraphics[scale=0.5]{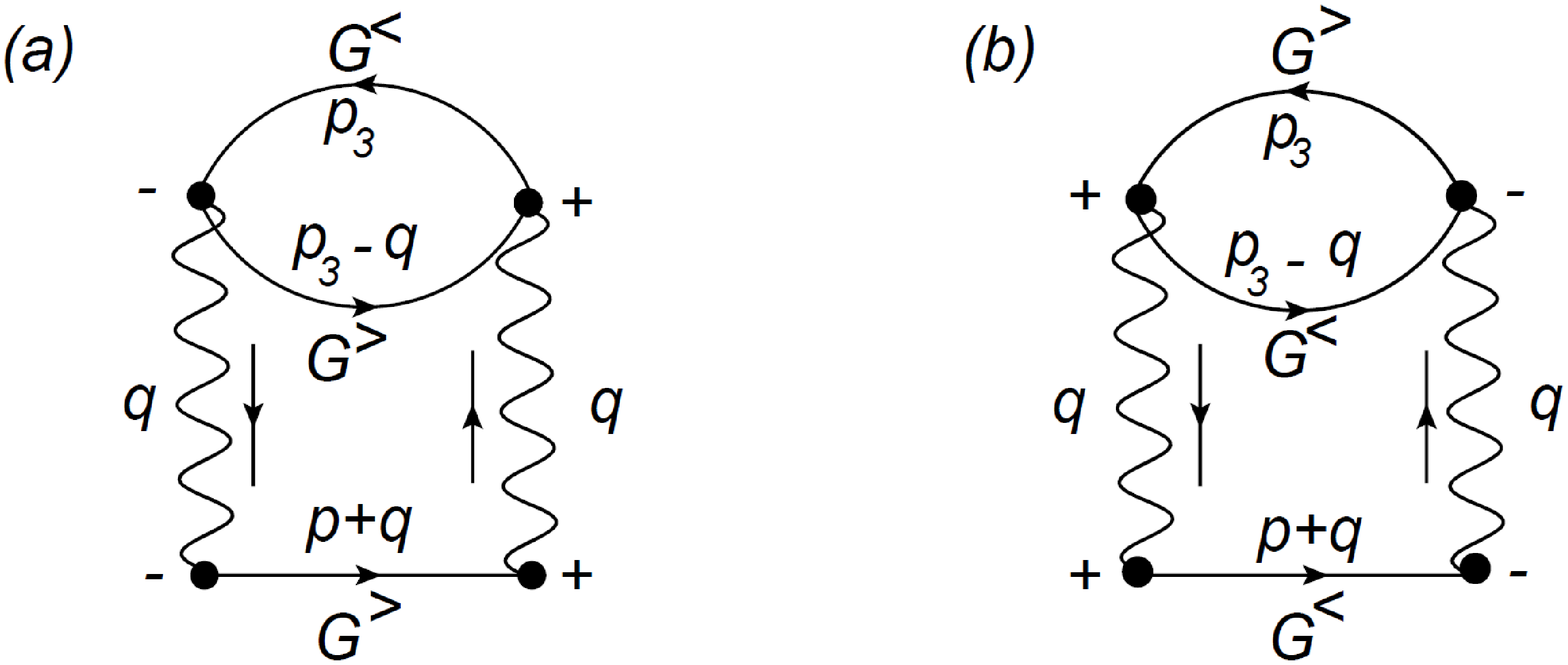}
\includegraphics[scale=0.5]{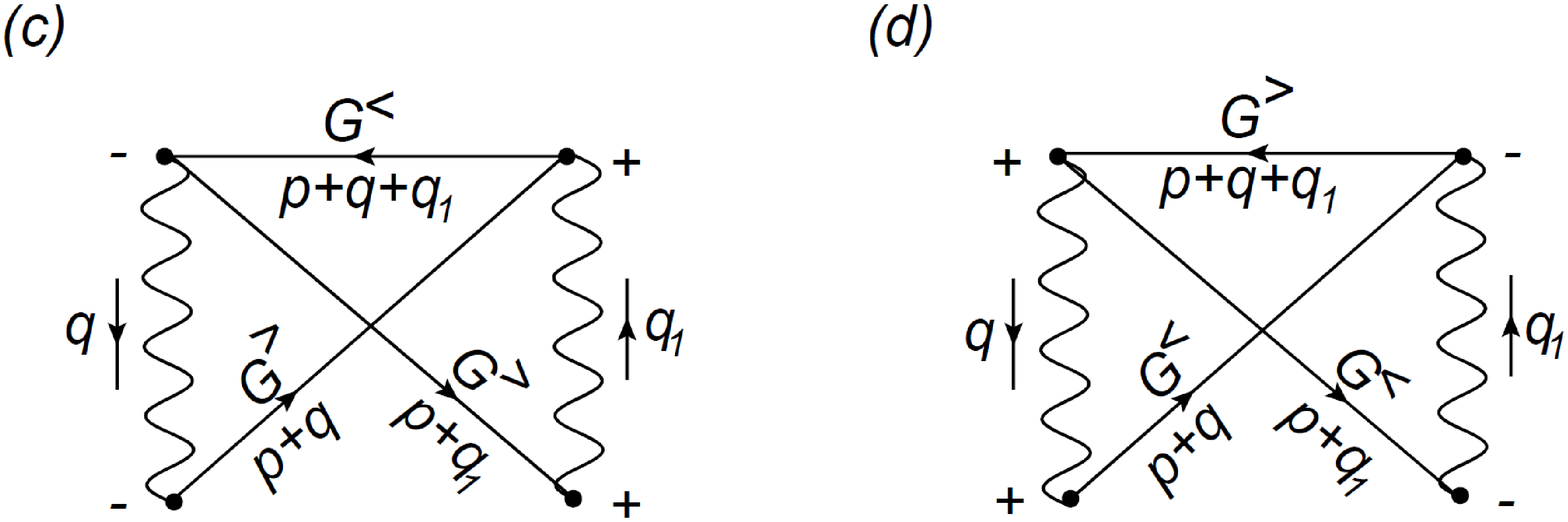}
\end{figure}

The collision term $I_{\mathrm{coll}}^{(1)}$ in Eq.\ (\ref{eq:collision-1st})
depends on the self-energies $\Sigma^{\lessgtr(0)}$.
Corresponding to the Feynman diagrams in Fig.\ \ref{fig:feynman}, we have
\begin{eqnarray}
\Sigma^{>}(x,p) & = & 4G_{a}G_{b}\int\frac{d^{4}p_{1}}{(2\pi\hbar)^{4}}
\frac{d^{4}p_{2}}{(2\pi\hbar)^{4}}\frac{d^{4}p_{3}}{(2\pi\hbar)^{4}}
(2\pi\hbar)^{4}\delta^{(4)}(p+p_{3}-p_{1}-p_{2})
\mathrm{Tr}\left[\Gamma_{a}G^{<}(x,p_{3})\Gamma_{b}G^{>}(x,p_{1})\right]
 \Gamma_{b}G^{>}(x,p_{2})\Gamma_{a}\nonumber \\
 &   -& 4G_{a}G_{b}\int\frac{d^{4}p_{1}}{(2\pi\hbar)^{4}}\frac{d^{4}p_{2}}{(2\pi\hbar)^{4}}
 \frac{d^{4}p_{3}}{(2\pi\hbar)^{4}}
 (2\pi\hbar)^{4}\delta^{(4)}(p+p_{3}-p_{1}-p_{2})
 \Gamma_{b}G^{>}(x,p_{1})\Gamma_{a}G^{<}(x,p_{3})\Gamma_{b}
 G^{>}(x,p_{2})\Gamma_{a}\;,\nonumber \\
\Sigma^{<}(x,p) & = & \Sigma^{>}[G^{>}\leftrightarrow G^{<}]\;.\label{eq:self-en-1}
\end{eqnarray}
The zeroth-order self-energies $\Sigma^{\lessgtr(0)}$ appearing
in $I_{\mathrm{coll}}^{(1)}$, Eq.\ (\ref{eq:collision-1st}), are
obtained from Eq.\ (\ref{eq:self-en-1}) by replacing $G^{\lessgtr}$
by their zeroth-order values as e.g.\ given in Eq.\ (\ref{eq:g-less-0}).
One can obtain $\Sigma^{<}$
from $\Sigma^{>}$ by interchanging $G^{>}\leftrightarrow G^{<}$ and
vice versa. Note that there is an overall sign difference between
the two terms in $\Sigma^{>}(x,p)$ {[}and similar for $\Sigma^{<}(x,p)${]}. The
first terms correspond to Figs.\ \ref{fig:feynman}(a) and (b),
while the second terms correspond to Figs.\ \ref{fig:feynman}(c) and (d), respectively.

We insert Eq.\ (\ref{eq:g-less-0}) for $G^{<(0)}$ and the corresponding
formula for $G^{>(0)}$ into Eq.\ (\ref{eq:self-en-1}) to obtain $\Sigma_{(0)}^{\lessgtr}$,
and then insert $\Sigma_{(0)}^{\lessgtr}$ and $G_{(0)}^{\lessgtr}$
into Eq.\ (\ref{eq:collision-1st}) to obtain $I_{\mathrm{coll}}^{(1)}$,
to finally obtain the Boltzmann equation (\ref{eq:boltzmann-fs-1st-order}) for
the scalar part in the form
\begin{eqnarray}
\lefteqn{\frac{1}{E_{p}}p\cdot\partial_{x}\mathrm{tr}\left[f^{(0)}(x,p)\right]
 =  \int_{0}^{\infty}\frac{dp_{0}}{2\pi\hbar}\,
\mathrm{Re}\mathrm{Tr}\left[\Sigma^{<(0)}(x,p)
G^{>(0)}(x,p)-\Sigma^{>(0)}(x,p)G^{<(0)}(x,p)\right]}\nonumber \\
 & = & \frac{1}{2E_{p}}\int\frac{d^{3}\mathbf{p}_{1}}{(2\pi\hbar)^{3}2E_{1}}
 \frac{d^{3}\mathbf{p}_{2}}{(2\pi\hbar)^{3}2E_{2}}
 \frac{d^{3}\mathbf{p}_{3}}{(2\pi\hbar)^{3}2E_{3}}
 (2\pi\hbar)^{4}\delta^{(4)}(p+p_{3}-p_{1}-p_{2})\nonumber \\
 &  & \times\mathrm{Re}\left\{\left[ f_1^{(0)} f_2^{(0)}\left(1-f_3^{(0)}\right)
 \left(1-f^{(0)}\right)-f_3^{(0)}f^{(0)} \left(1-f_1^{(0)}\right)
 \left(1-f_2^{(0)}\right)\right]  \left(M_{a}^{\mathrm{scalar}}
 +M_{b}^{\mathrm{scalar}}\right)\right\}\;,\label{eq:boltzman-fs-ns-scalar}
\end{eqnarray}
where $f_i^{(0)} \equiv f_{s_ir_i}^{(0)}(x,p_i)$,
$p_{i}^{\mu}=(E_{i},\mathbf{p}_{i}) \equiv( E_{p_{i}},\mathbf{p}_{i})$ ($i=1,2,3$)
and $f^{(0)} \equiv f_{sr}^{(0)}(x,p)$, $p^{\mu}=(E_{p},\mathbf{p})$,
and we denoted unit matrices in spin space as 1, e.g., $\delta _{rs} \rightarrow 1$.
The scalar spin-dependent matrix elements
$M_{a}^{\mathrm{scalar}}$ and $M_{b}^{\mathrm{scalar}}$ in the above
equation are defined as
\begin{eqnarray}
M_{a}^{\mathrm{scalar}} & \equiv & 4G_{a}G_{b}\mathrm{Tr}\left[\Gamma_{a}
u(r_{3},p_{3})\overline{u}(s_{3},p_{3})\Gamma_{b}u(r_{1},p_{1})
\overline{u}(s_{1},p_{1})\right] \mathrm{Tr}\left[\Gamma_{b}u(r_{2},p_{2})
 \overline{u}(s_{2},p_{2})\Gamma_{a}u(r,p)\overline{u}(s,p)\right],\nonumber \\
M_{b}^{\mathrm{scalar}} & \equiv & -4G_{a}G_{b}\mathrm{Tr}\left[\Gamma_{b}
u(r_{1},p_{1})\overline{u}(s_{1},p_{1})\Gamma_{a}u(r_{3},p_{3})
\overline{u}(s_{3},p_{3}) \Gamma_{b}u(r_{2},p_{2})\overline{u}(s_{2},p_{2})\Gamma_{a}
 u(r,p)\overline{u}(s,p)\right]\;.\label{eq:ma-mb-spin-dep}
\end{eqnarray}
Since these matrix elements depend on all spin indices,
sums over $s_{i},r_{i}$ ($i=1,2,3$) and $s,r$ are implied
on the right-hand side of Eq.\ (\ref{eq:boltzman-fs-ns-scalar}).

Following the same procedure as in the derivation of
Eq.\ (\ref{eq:boltzman-fs-ns-scalar}), the
Boltzmann equation (\ref{eq:boltzmann-fs-1st-order-1}) for the polarization
part can be written as
\begin{eqnarray}
\lefteqn{\hspace*{-1.7cm}
\frac{1}{E_{p}}p\cdot\partial_{x}\mathrm{tr}\left[n_{j}^{(+)\mu}\tau_{j}^{T}
f^{(0)}(x,p)\right]
 =  -\frac{1}{2m}\epsilon^{\mu\nu\alpha\beta}\int_{0}^{\infty}\frac{dp_{0}}{2\pi\hbar}
p_{\nu}\mathrm{Re}\mathrm{Tr}\left[\sigma_{\alpha\beta}\left(\Sigma^{<(0)}(x,p)
G^{>(0)}(x,p)-\Sigma^{>(0)}(x,p)G^{<(0)}(x,p)\right)\right]}\nonumber \\
 & = & -\frac{1}{4E_{p}m}\epsilon^{\mu\nu\alpha\beta}p_{\nu}
 \int\frac{d^{3}\mathbf{p}_{1}}{(2\pi\hbar)^{3}2E_{1}}
 \frac{d^{3}\mathbf{p}_{2}}{(2\pi\hbar)^{3}2E_{2}}
 \frac{d^{3}\mathbf{p}_{3}}{(2\pi\hbar)^{3}2E_{3}}
 (2\pi\hbar)^{4}\delta^{(4)}(p+p_{3}-p_{1}-p_{2})\nonumber \\
 &  & \times\mathrm{Re}\left\{\left[ f_1^{(0)} f_2^{(0)}\left(1-f_3^{(0)}\right)
 \left(1-f^{(0)}\right)-f_3^{(0)}f^{(0)} \left(1-f_1^{(0)}\right)
 \left(1-f_2^{(0)}\right)\right]
 \left(M_{a,\alpha\beta}^{\mathrm{pol}}+M_{b,\alpha\beta}^{\mathrm{pol}}\right)\right\}\;,
 \label{eq:boltzman-fs-ns-polar}
\end{eqnarray}
where the polarization parts of the spin-dependent matrix elements
$M_{a,\alpha\beta}^{\mathrm{pol}}$ and $M_{b,\alpha\beta}^{\mathrm{pol}}$
are defined as
\begin{eqnarray}
M_{a,\alpha\beta}^{\mathrm{pol}} & = & 4G_{a}G_{b}
\mathrm{Tr}\left[\Gamma_{a}u(r_{3},p_{3})\overline{u}(s_{3},p_{3})\Gamma_{b}
u(r_{1},p_{1})\overline{u}(s_{1},p_{1})\right]
\mathrm{Tr}\left[\sigma_{\alpha\beta}\Gamma_{b}u(r_{2},p_{2})\overline{u}(s_{2},p_{2})
\Gamma_{a}u(r,p)\overline{u}(s,p)\right]\;,\nonumber \\
M_{b,\alpha\beta}^{\mathrm{pol}} & = & -4G_{a}G_{b}
\mathrm{Tr}\left[\sigma_{\alpha\beta}\Gamma_{b}u(r_{1},p_{1})\overline{u}(s_{1},p_{1})
\Gamma_{a}u(r_{3},p_{3})\overline{u}(s_{3},p_{3})
\Gamma_{b}u(r_{2},p_{2})\overline{u}(s_{2},p_{2})\Gamma_{a}u(r,p)
\overline{u}(s,p)\right]\;.\label{eq:ma-mb-spin-dep-1}
\end{eqnarray}
An interesting observation is that, if the MVSD is
proportional to the unit matrix, i.e., only the first term in Eq.\ (\ref{eq:decomp_f0})
is nonzero, the left-hand side of Eq.\ (\ref{eq:boltzman-fs-ns-polar})
vanishes. Thus, also the collison term on the right-hand
side must vanish. In general, this is not obvious,
unless one is in global equilibrium, where already the term in brackets
vanishes on account of Eq.\ (\ref{eq:boltzman-fs-ns-scalar}).

Since we considered only fermions
in deriving Eq.\ (\ref{eq:boltzman-fs-ns-scalar}) from Eq.\ (\ref{eq:boltzmann-fs-1st-order}),
we have integrated over $p_{i0}$ ($i=1,2,3$) in the range $[0,+\infty]$.
For processes involving antifermions, we integrate over any
$p_{i0}$ ($i=1,2,3$) or $p_{0}$ in the range $[-\infty,0]$.
For an antifermion, the four-momentum
in the distribution function and the spinors becomes
$\overline{p}^{\mu}=(E_{p},-\mathbf{p})$.

In summary, we have shown in this section that
the MVSDs play the central role in the
Boltzmann equations for the scalar and polarization part,
which are derived from the corresponding equations for the scalar and
axial-vector components of the Wigner function. These equations are
closed and describe the time evolution of the MVSDs at
leading order in $\hbar$.

\section{Boltzmann equations with collisions at next-to-leading order}

\label{sec:boltzmann-next}

In this section we derive Boltzmann
equations at next-to-leading order $\mathcal{O}(\hbar^{2})$ from Eqs.\
(\ref{eq:boltzmann_scalar-2nd_v1}) and (\ref{eq:boltzmann_axialvector-2nd_v1}),
respectively. The
Boltzmann equation (\ref{eq:boltzmann_scalar-2nd_v1}) for $\mathcal{F}^{(1)}$
involves the collision terms
$I_{\mathrm{coll}}^{(2)}=\Delta I_{\mathrm{coll}}^{(1)}
+I_{\mathrm{coll},\mathrm{PB}}^{(0)}$
and $\partial_{x} I_{\mathrm{coll}}^{(1)}$, where the
collision term $\Delta I_{\mathrm{coll}}^{(1)}$, cf.\ Eq.\ (\ref{eq:def_DeltaI1}),
depends on the first-order Wigner functions $G^{\lessgtr(1)}$ as well
as the leading-order ones $G^{\lessgtr(0)}$,
while the collision term $I_{\mathrm{coll},\mathrm{PB}}^{(0)}$, cf.\ Eq.\ (\ref{eq:def_IPB}),
and $\partial_{x} I_{\mathrm{coll}}^{(1)}$ depend on $G^{\lessgtr(0)}$
only. The same holds for the Boltzmann equation
(\ref{eq:boltzmann_axialvector-2nd_v1})
for $\mathcal{A}^{(1)}_{\mu}$. Since these equations and the Wigner functions
are rather complicated at this order
[$\mathcal{O}(\hbar^{2})$ for the equations and $\mathcal{O}(\hbar)$ for
the Wigner functions], it is helpful to decompose the
full Wigner function at $\mathcal{O}(\hbar)$ into quasi-classical, gradient and
collision, as well as off-shell contributions, as shown in Eq.\ (\ref{eq:wig-kin-nonkin-off}).
The quasi-classical contributions (\ref{eq:1st-kin-comp_P})
-- (\ref{eq:1st-kin-comp_S}) and
the gradient and collision contributions (\ref{eq:1st-nonkin-comp_P})
-- (\ref{eq:1st-nonkin-comp_S})
are on-shell. The off-shell part $G_{\mathrm{off}}^{\lessgtr (1)}$, cf.\
Eqs.\ (\ref{eq:1st-off-comp_P}) -- (\ref{eq:1st-off-comp_S}),
will be neglected, since it is of higher order in the coupling constant as
discussed in Sec.\ \ref{sec:cancel-off}.

Now consider the collision term $\Delta I_{\mathrm{coll}}^{(1)}$
in Eq.\ (\ref{eq:def_DeltaI1}), which involves $\Sigma^{\lessgtr(1)}(x,p)$.
This function can be obtained from Eq.\ (\ref{eq:self-en-1}) by replacing
one $G^{\lessgtr}$ by $G^{\lessgtr(1)}$
and the other two by $G^{\lessgtr(0)}$. The quasi-classical
contributions $G^{\lessgtr(0)}$ and $G^{\lessgtr(1)}_\mathrm{qc}$
are given by Eqs.\ (\ref{eq:g-less-0}) and (\ref{eq:g-less-1}), respectively,
while the gradient and collision
contributions $G_{\nabla}^{\lessgtr(1)}$ can be
simplified as
\begin{eqnarray}
G_{\nabla}^{<(1)}=G_{\nabla}^{>(1)} & \approx &
-\frac{1}{8m}i\gamma^{5} \partial_{x}\cdot\mathcal{A}^{(0)}
-\frac{1}{8m}\gamma^{\nu}\partial_{x}^{\mu}\mathcal{S}_{\mu\nu}^{(0)}
+\frac{1}{8m}\sigma^{\mu\nu}\partial_{x,\mu}\mathcal{V}_{\nu}^{(0)}
 = \frac{i}{4m}\left[\gamma_\mu, \partial_{x}^{\mu}G^{<(0)}\right]\;.
 \label{eq:g-nonkinetic}
\end{eqnarray}
Here, we neglected terms that involve $I_{\mathrm{coll}}^{(1)}$,
since they are of higher order in the coupling constant
$G_c$, when we insert $G_{\nabla}^{\lessgtr(1)}$ into
$\Delta I_{\mathrm{coll}}^{(1)}$ on the right-hand side
of the Boltzmann equations.
We note that $G_{\nabla}^{\lessgtr(1)}$ contains only pseudoscalar,
vector, and tensor components as in Eqs.\ (\ref{eq:1st-nonkin-comp_P}) --
(\ref{eq:1st-nonkin-comp_S}).
Therefore we can write $\Delta I_{\mathrm{coll}}^{(1)}
=\Delta I_{\mathrm{coll,\,qc}}^{(1)}
+\Delta I_{\mathrm{coll,}\, \nabla}^{(1)}$
as the sum of quasi-classical and gradient plus collision contributions
$\Delta I_{\mathrm{coll,\,qc}}^{(1)}$
and $\Delta I_{\mathrm{coll,}\,\nabla}^{(1)}$, where Eq.\ (\ref{eq:g-nonkinetic})
is used for the first-order Wigner functions in
$\Delta I_{\mathrm{coll,}\, \nabla}^{(1)}$.

As an example for the first-order self-energy, we consider the first
term in Eq.\ (\ref{eq:self-en-1}), which corresponds
to Figs.\ \ref{fig:feynman} (a,b). This term
involves the following Dirac part in the integrand,
\begin{align}
&\mathrm{Tr}\left[\Gamma_{a}G^{<(1)}(x,p_{3})
\Gamma_{b}G^{>(0)}(x,p_{1})\right]\Gamma_{b}G^{>(0)}(x,p_{2})\Gamma_{a}
\nonumber \\
 +&\mathrm{Tr}\left[\Gamma_{a}G^{<(0)}(x,p_{3})\Gamma_{b}G^{>(1)}(x,p_{1})\right]
 \Gamma_{b}G^{>(0)}(x,p_{2})\Gamma_{a} \nonumber \\
 +& \mathrm{Tr}\left[\Gamma_{a}G^{<(0)}(x,p_{3})\Gamma_{b}G^{>(0)}(x,p_{1})\right]
 \Gamma_{b}G^{>(1)}(x,p_{2})\Gamma_{a}\;.\label{eq:sigma-1-ab}
\end{align}
On the other hand, the second term corresponding to Figs.\
\ref{fig:feynman} (c,d) involves the following Dirac part in the integrand,
\begin{align}
& \Gamma_{b}G^{>(1)}(x,p_{1})\Gamma_{a}G^{<(0)}(x,p_{3})
\Gamma_{b}G^{>(0)}(x,p_{2})\Gamma_{a} \nonumber \\
 + & \Gamma_{b}G^{>(0)}(x,p_{1})\Gamma_{a}G^{<(1)}(x,p_{3})
 \Gamma_{b}G^{>(0)}(x,p_{2})\Gamma_{a} \nonumber \\
 +& \Gamma_{b}G^{>(0)}(x,p_{1})\Gamma_{a}G^{<(0)}(x,p_{3})
 \Gamma_{b}G^{>(1)}(x,p_{2})\Gamma_{a}\;.\label{eq:sigma-1cd}
\end{align}
One can obtain $\Sigma^{<(1)}(x,p)$ from $\Sigma^{>(1)}(x,p)$
by exchanging $> \; \leftrightarrow\;  <$.
Note that $G^{\lessgtr(1)}$ in Eqs.\ (\ref{eq:sigma-1-ab}),
(\ref{eq:sigma-1cd}) are the full Wigner functions and
contain both quasi-classical as well as gradient and collision contributions.

Finally, we look at the nonlocal collision term
$I_{\mathrm{coll},\mathrm{PB}}^{(0)}$ defined in Eq.\ (\ref{eq:def_IPB}).
It is nonlocal due to the space-time derivatives appearing in the
Poisson brackets, so it represents collisions taking place at different
space-time points. We can write $I_{\mathrm{coll},\mathrm{PB}}^{(0)}$ explicitly
by expanding Poisson brackets
\begin{eqnarray}
I_{\mathrm{coll,PB}}^{(0)} & = & -\frac{1}{4}\left\{ \left[\partial_{x}^{\mu}\Sigma^{<(0)}(x,p)
\right] \left[ \partial_{\mu}^{p}G^{>(0)}(x,p)\right] -\left[ \partial_{p}^{\mu}\Sigma^{<(0)}(x,p)
\right] \left[ \partial_{\mu}^{x}G^{>(0)}(x,p)\right] \right.\nonumber \\
 &  & \left.-\left[ \partial_{x}^{\mu}\Sigma^{>(0)}(x,p)\right]
 \left[ \partial_{\mu}^{p}G^{<(0)}(x,p)\right] + \left[\partial_{p}^{\mu}\Sigma^{>(0)}(x,p)\right]
 \left[ \partial_{\mu}^{x}G^{<(0)}(x,p)\right]\right\} \;.\label{eq:i-coll-pb-1}
\end{eqnarray}
We observe that only the leading-order Wigner functions are relevant:
$I_{\mathrm{coll,PB}}^{(0)}$ is
fully determined once the leading-order MVSDs are known by solving
the Boltzmann equations (\ref{eq:boltzman-fs-ns-scalar}) and
(\ref{eq:boltzman-fs-ns-polar}).
The derivatives $\partial_{\mu}^{p}$ with respect to four-momentum have to be treated
carefully, since they act on delta functions.
There are two kinds of delta functions: one corresponds to energy-momentum
conservation in $\Sigma^{\lessgtr(0)}(x,p)$ and the other corresponds to on-shell
conditions in $G^{\lessgtr(0)}(x,p)$.
The derivatives of delta functions for energy-momentum conservation
are from $\partial_{p}^{\mu}\Sigma^{\lessgtr(0)}(x,p)$ and may be
connected with effects of energy non-conservation, while the derivatives
of mass-shell delta functions come from $\partial_{p_{0}}G^{\lessgtr(0)}(x,p)$
and give off-mass-shell contributions. A further and careful treatment of
four-momentum derivatives is essential for a numerical simulation of the spin
Boltzmann equations, but will not be made in this paper and will be addressed in
a future work.

Following the same procedure as in
the derivation of Eqs.\ (\ref{eq:boltzmann-fs-1st-order})
and (\ref{eq:boltzmann-fs-1st-order-1}), we obtain Boltzmann equations
at next-to-leading order from Eqs.\ (\ref{eq:boltzmann_scalar-2nd_v1})
and (\ref{eq:boltzmann_axialvector-2nd_v1}),
\begin{eqnarray}
\lefteqn{\hspace*{-2cm}\frac{1}{E_{p}}p\cdot\partial_{x}\mathrm{tr}\left[f^{(1)}(x,p)\right]  =
-\frac{1}{\pi\hbar}\int_{0}^{\infty}dp_{0}\,
\mathrm{Im}\mathrm{Tr}\left(I_{\mathrm{coll}}^{(2)}\right)
-\frac{1}{2\pi\hbar m}\, \mathrm{Re\,Tr}\left(\gamma\cdot\partial_{x}
I_{\mathrm{coll}}^{(1)}\right)} \nonumber  \\
 & \equiv & \mathscr{C}_{\mathrm{scalar}}\left(\Delta I_{\mathrm{coll},\,\mathrm{qc}}^{(1)}
 \right)+\mathscr{C}_{\mathrm{scalar}}\left(\Delta I_{\mathrm{coll,}\, \nabla}^{(1)}\right)
 +\mathscr{C}_{\mathrm{scalar}}\left(I_{\mathrm{coll},\mathrm{PB}}^{(0)}\right)
 +\mathscr{C}_{\mathrm{scalar}}\left(\partial_{x}I_{\mathrm{coll}}^{(1)}\right)\;,
 \label{eq:boltzmann-scalar-h2} \\
\lefteqn{\hspace*{-2cm}\frac{1}{E_{p}}p\cdot\partial_{x}\mathrm{tr}\left[n_{j}^{(+)\mu}
\tau_{j}^{T}f^{(1)}(x,p)\right]
 =  \frac{1}{2\pi\hbar m}\int_{0}^{\infty}dp_{0}\, \left[\epsilon^{\mu\nu\alpha\beta}p_{\nu}
\mathrm{Im}\mathrm{Tr}\left(\sigma_{\alpha\beta}I_{\mathrm{coll}}^{(2)}\right)
+\mathrm{Re\,Tr}\left(\gamma^{5}\partial_{x}^{\mu}I_{\mathrm{coll}}^{(1)}\right)\right]}
\nonumber \\
 & \equiv & \mathscr{C}^\mu_{\mathrm{pol}}
 \left(\Delta I_{\mathrm{coll},\,\mathrm{qc}}^{(1)}
 \right)+\mathscr{C}^\mu_{\mathrm{pol}}\left(\Delta I_{\mathrm{coll,}\, \nabla}^{(1)}\right)
 +\mathscr{C}^\mu_{\mathrm{pol}}\left(I_{\mathrm{coll},\mathrm{PB}}^{(0)}\right)
 +\mathscr{C}^\mu_{\mathrm{pol}}\left(\partial_{x}I_{\mathrm{coll}}^{(1)}\right)\;.
 \label{eq:boltzmann-axial-h2}
\end{eqnarray}
The first Boltzmann equation is for the scalar part of the first-order MVSD, while
the second one is for the polarization part of the first-order MVSD. Due to the fact
that $I_{\mathrm{coll}}^{(2)}$ consists of $\Delta I_{\mathrm{coll}}^{(1)}$
and $I_{\mathrm{coll},\mathrm{PB}}^{(0)}$, we define the collision terms
on the right-hand sides of Eqs.\ (\ref{eq:boltzmann-scalar-h2}),
(\ref{eq:boltzmann-axial-h2})
according to their origins. Note that the
collision terms involving $\Delta I_{\mathrm{coll},\,\mathrm{qc}}^{(1)}$
depend on $f^{(1)}(x,p)$ as well as on $f^{(0)}(x,p)$ in an
algebraic way, while those involving $\Delta I_{\mathrm{coll,}\, \nabla}^{(1)}$,
$I_{\mathrm{coll},\mathrm{PB}}^{(0)}$ and $\partial_{x}I_{\mathrm{coll}}^{(1)}$
contain space-time derivatives of $f^{(0)}(x,p)$, which are determined
by solving the leading-order Boltzmann equations (\ref{eq:boltzman-fs-ns-scalar}),
(\ref{eq:boltzman-fs-ns-polar}). Therefore, the collision terms involving
$\Delta I_{\mathrm{coll,}\, \nabla}^{(1)}$, $I_{\mathrm{coll},\mathrm{PB}}^{(0)}$,
and $\partial_{x}I_{\mathrm{coll}}^{(1)}$ containing space-time derivatives
of $f^{(0)}(x,p)$ can be regarded as source terms for the scalar and
polarization parts of $f^{(1)}(x,p)$.

The explicit form of the local collision term
$\mathscr{C}_{\mathrm{scalar}}\left(\Delta I_{\mathrm{coll},\,\mathrm{qc}}^{(1)}\right)$
is given by
\begin{eqnarray}
\mathscr{C}_{\mathrm{scalar}}\left(\Delta I_{\mathrm{coll},\,\mathrm{qc}}^{(1)}\right)
& = & -\frac{1}{\pi\hbar}\int_{0}^{\infty}dp_{0}\,
\mathrm{Im\,Tr}\left(\Delta I_{\mathrm{coll},\,\mathrm{qc}}^{(1)}\right)\nonumber \\
 & = & \frac{1}{2E_{p}}\int\frac{d^{3}\mathbf{p}_{1}}{(2\pi\hbar)^{3}2E_{1}}
 \frac{d^{3}\mathbf{p}_{2}}{(2\pi\hbar)^{3}2E_{2}}
 \frac{d^{3}\mathbf{p}_{3}}{(2\pi\hbar)^{3}2E_{3}}\,
 (2\pi\hbar)^{4}\delta^{(4)}(p+p_{3}-p_{1}-p_{2})\nonumber \\
 &  & \times\mathrm{Re}\left[\left\{ f^{(1)}_{1}\left[f^{(0)}_{2}\left(1-f^{(0)}-f^{(0)}_{3}\right)
 +f^{(0)}_{3}f^{(0)}\right]
 +f^{(1)}_{2}\left[f^{(0)}_{1}\left(1-f^{(0)}_{3}-f^{(0)}\right)+f^{(0)}_{3}f^{(0)}\right]\right.\right.
 \nonumber \\
 &  & \hspace*{0.8cm}
  -f^{(1)}\left[f^{(0)}_{3}\left(1-f^{(0)}_{1}-f^{(0)}_{2}\right)+f^{(0)}_{1}f^{(0)}_{2}\right]
  \left.-f^{(1)}_{3}\left[f^{(0)}\left(1-f^{(0)}_{1}-f^{(0)}_{2}\right)+f^{(0)}_{1}f^{(0)}_{2}
 \right]\right\} \nonumber \\
 &  & \hspace*{0.7cm}
 \times\left.\left(M_{a}^{\mathrm{scalar}}+M_{b}^{\mathrm{scalar}}\right)\right]\;,
 \label{eq:coll-scalar-1}
\end{eqnarray}
where similar to the notation employed above for $f^{(0)}_i$ we
have defined
$f^{(1)}_i \equiv f_{s_{i}r_{i}}^{(1)}(x,p_{i})$ ($i=1,2,3$)
and $f^{(1)}\equiv f_{sr}^{(1)}(x,p)$. The scalar parts of the
matrix elements $M_{a,b}^{\mathrm{scalar}}$ are given by Eq.\ (\ref{eq:ma-mb-spin-dep}).
Note that
$\mathscr{C}_{\mathrm{scalar}}\left(\Delta I_{\mathrm{coll},\,\mathrm{qc}}^{(1)}\right)$
is actually the first-order perturbation to the collision term of
the leading-order Boltzmann equation (\ref{eq:boltzman-fs-ns-scalar}).
The explicit form of the local collision term
$\mathscr{C}_{\mathrm{pol}}\left(\Delta I_{\mathrm{coll},\,\mathrm{qc}}^{(1)}\right)$
is given by
\begin{eqnarray}
\mathscr{C}^\mu_{\mathrm{pol}}\left(\Delta I_{\mathrm{coll},\,\mathrm{qc}}^{(1)}\right)
& = & \frac{1}{2\pi\hbar m}\int_{0}^{\infty}dp_{0}\, \epsilon^{\mu\nu\alpha\beta}p_{\nu}
\mathrm{Im}\mathrm{Tr}\left(\sigma_{\alpha\beta}
\Delta I_{\mathrm{coll},\,\mathrm{qc}}^{(1)}\right)\nonumber \\
 & = & -\frac{1}{4E_{p}m}\,\epsilon^{\mu\nu\alpha\beta}p_{\nu}
 \int\frac{d^{3}\mathbf{p}_{1}}{(2\pi\hbar)^{3}2E_{1}}
 \frac{d^{3}\mathbf{p}_{2}}{(2\pi\hbar)^{3}2E_{2}}
 \frac{d^{3}\mathbf{p}_{3}}{(2\pi\hbar)^{3}2E_{3}}\,
 (2\pi\hbar)^{4}\delta^{(4)}(p+p_{3}-p_{1}-p_{2})\nonumber \\
 &  & \times\mathrm{Re}\left[\left\{ f^{(1)}_{1}\left[f^{(0)}_{2}\left(1-f^{(0)}-f^{(0)}_{3}\right)
 +f^{(0)}_{3}f^{(0)}\right]
 +f^{(1)}_{2}\left[f^{(0)}_{1}\left(1-f^{(0)}_{3}-f^{(0)}\right)+f^{(0)}_{3}f^{(0)}\right]\right.\right.
 \nonumber \\
 &  & \hspace*{0.8cm}
  -f^{(1)}\left[f^{(0)}_{3}\left(1-f^{(0)}_{1}-f^{(0)}_{2}\right)+f^{(0)}_{1}f^{(0)}_{2}\right]
  \left.-f^{(1)}_{3}\left[f^{(0)}\left(1-f^{(0)}_{1}-f^{(0)}_{2}\right)+f^{(0)}_{1}f^{(0)}_{2}
 \right]\right\} \nonumber \\
 &  & \hspace*{0.7cm} \times\left.\left(M_{a,\alpha\beta}^{\mathrm{pol}}
 +M_{b,\alpha\beta}^{\mathrm{pol}}\right)\right]\;, \label{eq:c-pol-delta-I}
\end{eqnarray}
where the polarization parts of the
matrix elements $M_{a,b;\alpha\beta}^{\mathrm{pol}}$
are given by Eq.\ (\ref{eq:ma-mb-spin-dep-1}).
The polarization part of the collision term
$\mathscr{C}^\mu_{\mathrm{pol}}\left(\Delta I_{\mathrm{coll,\, qc}}^{(1)}\right)$
is actually the first-order perturbation to the collision term of
the leading-order Boltzmann equation (\ref{eq:boltzman-fs-ns-polar}).
From Eqs.\ (\ref{eq:coll-scalar-1}), (\ref{eq:c-pol-delta-I})
we observe
that the quasi-classical parts of the collision terms depend
on $f^{(1)}$ as well as on $f^{(0)}$ and do not contain space-time derivatives.

The gradient parts of the collision terms are defined as
\begin{eqnarray}
\mathscr{C}_{\mathrm{scalar}}\left(\Delta I_{\mathrm{coll,}\, \nabla}^{(1)}\right)
& = & -\frac{1}{\pi\hbar m}\int_{0}^{\infty}dp_{0}\,
\mathrm{Im\,Tr}\left(p\cdot\gamma\Delta I_{\mathrm{coll,}\, \nabla}^{(1)}\right)\;,
\nonumber \\
\mathscr{C}^\mu_{\mathrm{pol}}\left(\Delta I_{\mathrm{coll,}\, \nabla}^{(1)}\right)
& = & \frac{1}{2\pi\hbar m}\int_{0}^{\infty}dp_{0}\epsilon^{\mu\nu\alpha\beta}p_{\nu}
\mathrm{Im}\mathrm{Tr}\left(\sigma_{\alpha\beta}
\Delta I_{\mathrm{coll},\,\nabla}^{(1)}\right)\;,
\end{eqnarray}
which are nonlocal since they involve space-time derivatives of $f^{(0)}$
as shown in Eq.\ (\ref{eq:g-nonkinetic}) for $G_{\nabla}^{\lessgtr(1)}$.

The explicit form of the nonlocal collision term
$\mathscr{C}_{\mathrm{scalar}}\left(\partial_x I_{\mathrm{coll}}^{(1)}\right)$
reads
\begin{eqnarray*}
\mathscr{C}_{\mathrm{scalar}}\left(\partial_{x}I_{\mathrm{coll}}^{(1)}\right)
& = & -\frac{1}{2\pi\hbar m}\mathrm{Re\,Tr}\left(\gamma\cdot\partial_{x}
I_{\mathrm{coll}}^{(1)}\right)\\
 & = & -\frac{1}{4mE_{p}}\int\frac{d^{3}\mathbf{p}_{1}}{(2\pi\hbar)^{3}2E_{1}}
 \frac{d^{3}\mathbf{p}_{2}}{(2\pi\hbar)^{3}2E_{2}}
 \frac{d^{3}\mathbf{p}_{3}}{(2\pi\hbar)^{3}2E_{3}}\,
 (2\pi\hbar)^{4}\delta^{(4)}(p+p_{3}-p_{1}-p_{2})\\
 &  & \times\mathrm{Im}\left\{\partial_{x}^{\mu}\left[ f^{(0)}_1f^{(0)}_2
 \left(1-f^{(0)}_3\right) \left(1-f^{(0)}\right)
 -f^{(0)}_3f^{(0)}\left(1-f^{(0)}_{1}\right)\left(1-f^{(0)}_2\right)\right]
 \left(M_{a,\mu}^{\mathrm{scalar}}+M_{b,\mu}^{\mathrm{scalar}}\right)\right\}\;,
\end{eqnarray*}
where the matrix elements
$M_{a,\mu}^{\mathrm{scalar}},\, M_{b,\mu}^{\mathrm{scalar}}$ are defined by
\begin{eqnarray}
M_{a,\mu}^{\mathrm{scalar}} & = &
4G_{a}G_{b}\mathrm{Tr}\left[\Gamma_{a}u(r_{3},p_{3})
\overline{u}(s_{3},p_{3})\Gamma_{b}u(r_{1},p_{1})\overline{u}(s_{1},p_{1})\right]
 \mathrm{Tr}\left[\gamma_{\mu}\Gamma_{b}u(r_{2},p_{2})
 \overline{u}(s_{2},p_{2})\Gamma_{a}u(r,p)\overline{u}(s,p)\right]\;,\nonumber \\
M_{b,\mu}^{\mathrm{scalar}} & = & -4G_{a}G_{b}\mathrm{Tr}\left[\gamma_{\mu}
\Gamma_{b}u(r_{1},p_{1})\overline{u}(s_{1},p_{1})\Gamma_{a}u(r_{3},p_{3})
\overline{u}(s_{3},p_{3}) \Gamma_{b}u(r_{2},p_{2})\overline{u}(s_{2},p_{2})\Gamma_{a}
 u(r,p)\overline{u}(s,p)\right]\;.\label{eq:matrix-element-spin-1}
\end{eqnarray}
The explicit form of the nonlocal collision term
$\mathscr{C}^\mu_{\mathrm{pol}}\left(\partial_x I_{\mathrm{coll}}^{(1)}\right)$ reads
\begin{eqnarray}
\lefteqn{\mathscr{C}^\mu_{\mathrm{pol}}\left(\partial_{x}I_{\mathrm{coll}}^{(1)}\right)
 =  \frac{1}{2\pi\hbar m}\int_{0}^{\infty}dp_{0}\, \mathrm{Re\,Tr}
\left(\gamma^{5}\partial_{x}^{\mu}I_{\mathrm{coll}}^{(1)}\right)}\nonumber \\
 & = & \frac{1}{4mE_{p}}\int\frac{d^{3}\mathbf{p}_{1}}{(2\pi\hbar)^{3}2E_{1}}
 \frac{d^{3}\mathbf{p}_{2}}{(2\pi\hbar)^{3}2E_{2}}
 \frac{d^{3}\mathbf{p}_{3}}{(2\pi\hbar)^{3}2E_{3}}\,
 (2\pi\hbar)^{4}\delta^{(4)}(p+p_{3}-p_{1}-p_{2})\nonumber \\
 &  & \times\mathrm{Im}\left\{\partial_{x}^{\mu}\left[ f^{(0)}_1f^{(0)}_2\left(1-f^{(0)}_3\right)\
 \left(1-f^{(0)}\right)
 -f^{(0)}_3 f^{(0)}\left(1-f^{(0)}_1\right)\left(1-f^{(0)}_2\right)\right]
\left(M_{a,5}^{\mathrm{pol}}+M_{b,5}^{\mathrm{pol}}\right)\right\}\;,
\end{eqnarray}
where the matrix elements
$M_{a,5}^{\mathrm{pol}},\, M_{b,5}^{\mathrm{pol}}$ are defined by
\begin{eqnarray}
M_{a,5}^{\mathrm{pol}} & = & 4G_{a}G_{b}\mathrm{Tr}\left[\Gamma_{a}u(r_{3},p_{3})
\overline{u}(s_{3},p_{3})\Gamma_{b}u(r_{1},p_{1})\overline{u}(s_{1},p_{1})\right]
\mathrm{Tr}\left[\gamma_{5}
 \Gamma_{b}u(r_{2},p_{2})\overline{u}(s_{2},p_{2})\Gamma_{a}u(r,p)
 \overline{u}(s,p)\right]\,,\nonumber \\
M_{b,5}^{\mathrm{pol}} & = & -4G_{a}G_{b}\mathrm{Tr}\left[\gamma_{5}
\Gamma_{b}u(r_{1},p_{1})\overline{u}(s_{1},p_{1})
\Gamma_{a}u(r_{3},p_{3})\overline{u}(s_{3},p_{3})
\Gamma_{b}u(r_{2},p_{2})\overline{u}(s_{2},p_{2})\Gamma_{a}u(r,p)\overline{u}(s,p)\right]
\;.\label{eq:matrix-element-spin-2}
\end{eqnarray}

The nonlocal collision terms involving Poisson brackets are
\begin{eqnarray}
\mathscr{C}_{\mathrm{scalar}}\left( I_{\mathrm{coll},\mathrm{PB}}^{(0)} \right)
& = &  -\frac{1}{\pi\hbar}\int_{0}^{\infty}dp_{0}\,
\mathrm{Im}\mathrm{Tr}\left(I_{\mathrm{coll},\mathrm{PB}}^{(0)}\right)
\,,\nonumber\\
\mathscr{C}^\mu_{\mathrm{pol}}\left(I_{\mathrm{coll},\mathrm{PB}}^{(0)}\right)
& = & \frac{1}{2\pi\hbar m}\int_{0}^{\infty}dp_{0}\, \epsilon^{\mu\nu\alpha\beta}p_{\nu}
\mathrm{Im}\mathrm{Tr}\left( \sigma_{\alpha\beta}I_{\mathrm{coll},\mathrm{PB}}^{(0)}
\right) \,,
\end{eqnarray}
where $I_{\mathrm{coll},\mathrm{PB}}^{(0)}$ is given by Eq.\ (\ref{eq:i-coll-pb-1}).
We refrain from giving the explicit forms of the above collision terms, as they
are too lengthy.

We conclude this section with some remarks about the results: (a) For the
gradient and collision contributions of the Wigner functions, we assume
Eq.\ (\ref{eq:g-nonkinetic}), i.e.,
the scalar component $\mathcal{F}_{\nabla}^{(1)}$ and the
axial-vector component $\mathcal{A}_{\nabla}^{(1)\mu}$ are set
to zero. This is the most general solution to $G_{\nabla}^{\lessgtr(1)}$
that satisfies Eqs.\ (\ref{eq:sol-p1}) -- (\ref{eq:sol-a1}).
If $\mathcal{F}_{\nabla}^{(1)}$
and $\mathcal{A}_{\nabla}^{(1)\mu}$ are nonvanishing, it
can be proved that they can be absorbed into the kinetic
contributions of the Wigner functions by a redefinition of the first-order MVSD $f^{(1)}$.
(b) The set of Boltzmann equations (\ref{eq:boltzmann-scalar-h2})
and (\ref{eq:boltzmann-axial-h2}) for the
MVSD $f^{(1)}$ at $\mathcal{O}(\hbar)$
can be solved once the
Boltzmann equations (\ref{eq:boltzman-fs-ns-scalar})
and (\ref{eq:boltzman-fs-ns-polar}) have been solved
for $f^{(0)}$ at $\mathcal{O}(\hbar^{0})$.
Once $f^{(0)}$ is known, the nonlocal terms arising from space-time
derivatives of $f^{(0)}$ play the role of source terms for
the collisions terms in the
Boltzmann equations (\ref{eq:boltzmann-scalar-h2})
and (\ref{eq:boltzmann-axial-h2}). Therefore, the polarization part
of $f^{(1)}$ is driven by space-time derivatives of $f^{(0)}$,
which are proportional to the thermal vorticity in global equilibrium.

\section{Summary}
\label{sec:summary}

The closed-time-path formalism is an effective method to deal with
non-equilibrium problems. In this paper, we derived
the Kadanoff--Baym equation from
the Dyson-Schwinger equation on the CTP contour for massive spin-1/2
fermions, where collisions are provided by the
irreducible self-energy. We focussed
on the Kadanoff--Baym equation for the two-point function $G^{<}(x_{1},x_{2})$,
which is a $4\times4$ matrix in Dirac space. We performed a
gradient expansion
for the self-energy and took the Fourier transform of the resulting
equation with respect to the distance between two space-time points.
Thus we derived the Kadanoff--Baym equation for the Wigner function
$G^{<}(x,p)$. The self-energy is expressed in terms of the Wigner
functions $G^{<}(x,p)$ and $G^{>}(x,p)$. We employed a semi-classical
expansion of the Kadanoff--Baym equation for the two-point function
in powers of $\hbar$. By projecting the matrix form of the
Kadanoff--Baym equation onto the Dirac matrices of the Clifford decomposition,
we derived a set of equations in terms of the Clifford components of the
Wigner function up to $\hbar^{2}$. The scalar component of the Wigner
function corresponds to the phase-space
distribution function, while the axial-vector component carries the
information for the phase-space distribution of the spin polarization.

At leading order $\mathcal{O}(\hbar ^0)$, the Wigner functions
$G^{<(0)}(x,p)$ and $G^{>(0)}(x,p)$
can be expressed in terms of matrix-valued spin distribution functions (MVSDs).
The form of $G^{\lessgtr(0)}(x,p)$
is the same as in kinetic theory without collisions.
At $\mathcal{O}(\hbar)$ or next-to-leading order, the
Wigner functions $G^{<(1)}(x,p)$ and $G^{>(1)}(x,p)$
can be separated into (on-shell) quasi-classical, (on-shell)
gradient and collision, as well as off-shell contributions.
The quasi-classical contributions to $G^{<(1)}(x,p)$ and $G^{>(1)}(x,p)$
can be obtained by considering the first-order correction
$f^{(1)}(x,p)$ to the leading-order MVSD $f^{(0)}(x,p)$.
The gradient and collision contributions contain, besides collision terms,
space-time derivatives of zeroth-order Wigner functions in the equations of motions.
The off-shell contributions are the ones that violate the on-shell conditions.
These belong to higher-order contributions in the coupling constant
and can be decoupled from the on-shell parts of the Boltzmann equations.
The Boltzmann equations for the scalar and axial-vector (i.e., polarization) parts
can be expressed in terms of the MVSDs. At $\mathcal{O}(\hbar)$ or leading order,
only local collision terms appear in the Boltzmann equations, without
space-time derivatives, meaning that collisions take place at the same space-time
point. At $\mathcal{O}(\hbar^2)$ or next-to-leading order, the
Boltzmann equations describe how $f^{(1)}(x,p)$
evolves under the influence of local as well as nonlocal collision
terms with space-time derivatives.
Nonlocal collision terms depend on the leading-order
MVSD $f^{(0)}(x,p)$ and its space-time derivative
(generating the vorticity in equilibrium), which are determined by solving
the leading-order Boltzmann equations,
while the local collision terms depend on $f^{(1)}(x,p)$
as well as $f^{(0)}(x,p)$.
Therefore, the nonlocal collision terms can be regarded as sources for
the polarization part of $f^{(1)}(x,p)$ in the
Boltzmann equations at next-to-leading order.
The system of Boltzmann equations in terms of the MVSDs paves the way for simulating
spin transport processes from first principles.


\begin{acknowledgments}
The authors thank F.\ Becattini, W.\ Florkowski, X.\ Guo, U.\ Heinz, Y.-C.\
Liu, R.\ Ryblewski, L.\ Tinti, and G.\ Torrieri for enlightening
discussions. Q.W.\ is supported in part by the National
Natural Science Foundation of China (NSFC) under Grant No.\
11890713 (a sub-grant of 11890710),
and 11947301, and by the Strategic Priority Research Program of Chinese
Academy of Sciences under Grant No.\ XDB34030102.
X.L.S.\ is supported in part by the National
Natural Science Foundation of China (NSFC) under Grant No.\ 12047528 and 11890714.
The work of D.H.R., E.S., and N.W.\ is supported by the
Deutsche Forschungsgemeinschaft
(DFG, German Research Foundation) through the Collaborative Research
Center CRC-TR 211 ``Strong-interaction matter under extreme conditions''
- project number 315477589 - TRR 211. E.S.\ acknowledges support by
BMBF ``Forschungsprojekt: 05P2018 - Ausbau von ALICE am LHC (05P18RFCA1)''.
\end{acknowledgments}

\bibliographystyle{unsrt}
\bibliography{kb-paper-ref}

\begin{thebibliography}{10}

\bibitem{Rischke:2003mt}
Dirk~H. Rischke.
\newblock {The Quark gluon plasma in equilibrium}.
\newblock {\em Prog. Part. Nucl. Phys.}, 52:197--296, 2004.

\bibitem{Gyulassy:2004zy}
Miklos Gyulassy and Larry McLerran.
\newblock {New forms of QCD matter discovered at RHIC}.
\newblock {\em Nucl. Phys. A}, 750:30--63, 2005.

\bibitem{Shuryak:2004cy}
Edward~V. Shuryak.
\newblock {What RHIC experiments and theory tell us about properties of
  quark-gluon plasma?}
\newblock {\em Nucl. Phys. A}, 750:64--83, 2005.

\bibitem{Ackermann:2000tr}
K.H. Ackermann et~al.
\newblock {Elliptic flow in Au + Au collisions at (S(NN))**(1/2) = 130 GeV}.
\newblock {\em Phys. Rev. Lett.}, 86:402--407, 2001.

\bibitem{Kovtun:2004de}
P.~Kovtun, Dan~T. Son, and Andrei~O. Starinets.
\newblock {Viscosity in strongly interacting quantum field theories from black
  hole physics}.
\newblock {\em Phys. Rev. Lett.}, 94:111601, 2005.

\bibitem{Kolb:2003dz}
Peter~F. Kolb and Ulrich~W. Heinz.
\newblock {Hydrodynamic description of ultrarelativistic heavy ion collisions}.
\newblock pages 634--714, 5 2003.

\bibitem{Heinz:2013th}
Ulrich Heinz and Raimond Snellings.
\newblock {Collective flow and viscosity in relativistic heavy-ion collisions}.
\newblock {\em Ann. Rev. Nucl. Part. Sci.}, 63:123--151, 2013.

\bibitem{Florkowski:2017olj}
Wojciech Florkowski, Michal~P. Heller, and Michal Spalinski.
\newblock {New theories of relativistic hydrodynamics in the LHC era}.
\newblock {\em Rept. Prog. Phys.}, 81(4):046001, 2018.

\bibitem{Romatschke:2017ejr}
Paul Romatschke and Ulrike Romatschke.
\newblock {\em {Relativistic Fluid Dynamics In and Out of Equilibrium}}.
\newblock Cambridge Monographs on Mathematical Physics. Cambridge University
  Press, 5 2019.

\bibitem{Barnett:1935}
S.~J. Barnett.
\newblock Gyromagnetic and electron-inertia effects.
\newblock {\em Rev. Mod. Phys.}, 7:129--166, Apr 1935.

\bibitem{Liang:2004ph}
Zuo-Tang Liang and Xin-Nian Wang.
\newblock {Globally polarized quark-gluon plasma in non-central A+A
  collisions}.
\newblock {\em Phys. Rev. Lett.}, 94:102301, 2005.
\newblock [Erratum: Phys. Rev. Lett.96,039901(2006)].

\bibitem{Liang:2004xn}
Zuo-Tang Liang and Xin-Nian Wang.
\newblock {Spin alignment of vector mesons in non-central A+A collisions}.
\newblock {\em Phys. Lett. B}, 629:20--26, 2005.

\bibitem{Voloshin:2004ha}
Sergei~A. Voloshin.
\newblock {Polarized secondary particles in unpolarized high energy
  hadron-hadron collisions?}
\newblock 10 2004.

\bibitem{Betz:2007kg}
Barbara Betz, Miklos Gyulassy, and Giorgio Torrieri.
\newblock {Polarization probes of vorticity in heavy ion collisions}.
\newblock {\em Phys. Rev. C}, 76:044901, 2007.

\bibitem{Becattini:2007sr}
F.~Becattini, F.~Piccinini, and J.~Rizzo.
\newblock {Angular momentum conservation in heavy ion collisions at very high
  energy}.
\newblock {\em Phys. Rev. C}, 77:024906, 2008.

\bibitem{Abelev:2007zk}
B.I. Abelev et~al.
\newblock {Global polarization measurement in Au+Au collisions}.
\newblock {\em Phys. Rev. C}, 76:024915, 2007.
\newblock [Erratum: Phys.Rev.C 95, 039906 (2017)].

\bibitem{STAR:2017ckg}
L.~Adamczyk et~al.
\newblock {Global $\Lambda$ hyperon polarization in nuclear collisions:
  evidence for the most vortical fluid}.
\newblock {\em Nature}, 548:62--65, 2017.

\bibitem{Adam:2018ivw}
Jaroslav Adam et~al.
\newblock {Global polarization of $\Lambda$ hyperons in Au+Au collisions at
  $\sqrt{s_{_{NN}}}$ = 200 GeV}.
\newblock {\em Phys. Rev.}, C98:014910, 2018.

\bibitem{Wang:2017jpl}
Qun Wang.
\newblock {Global and local spin polarization in heavy ion collisions: a brief
  overview}.
\newblock {\em Nucl. Phys. A}, 967:225--232, 2017.

\bibitem{Becattini:2020ngo}
Francesco Becattini and Michael~A. Lisa.
\newblock {Polarization and Vorticity in the Quark Gluon Plasma}.
\newblock 3 2020.

\bibitem{Gao:2020vbh}
Jian-Hua Gao, Guo-Liang Ma, Shi Pu, and Qun Wang.
\newblock {Recent developments in chiral and spin polarization effects in heavy
  ion collisions}.
\newblock 5 2020.

\bibitem{Karpenko:2016jyx}
I.~Karpenko and F.~Becattini.
\newblock {Study of $\Lambda $ polarization in relativistic nuclear collisions
  at $\sqrt{s_\mathrm {NN}}=7.7$ --200 GeV}.
\newblock {\em Eur. Phys. J. C}, 77(4):213, 2017.

\bibitem{Xie:2017upb}
Yilong Xie, Dujuan Wang, and Laszlo~P. Csernai.
\newblock {Global $\Lambda$ polarization in high energy collisions}.
\newblock {\em Phys. Rev. C}, 95(3):031901, 2017.

\bibitem{Li:2017slc}
Hui Li, Long-Gang Pang, Qun Wang, and Xiao-Liang Xia.
\newblock {Global $\Lambda$ polarization in heavy-ion collisions from a
  transport model}.
\newblock {\em Phys. Rev. C}, 96(5):054908, 2017.

\bibitem{Sun:2017xhx}
Yifeng Sun and Che~Ming Ko.
\newblock {$\Lambda$ hyperon polarization in relativistic heavy ion collisions
  from a chiral kinetic approach}.
\newblock {\em Phys. Rev. C}, 96(2):024906, 2017.

\bibitem{Wei:2018zfb}
De-Xian Wei, Wei-Tian Deng, and Xu-Guang Huang.
\newblock {Thermal vorticity and spin polarization in heavy-ion collisions}.
\newblock {\em Phys. Rev. C}, 99(1):014905, 2019.

\bibitem{Baznat:2013zx}
Mircea Baznat, Konstantin Gudima, Alexander Sorin, and Oleg Teryaev.
\newblock {Helicity separation in Heavy-Ion Collisions}.
\newblock {\em Phys. Rev. C}, 88(6):061901, 2013.

\bibitem{Csernai:2013bqa}
L.P. Csernai, V.K. Magas, and D.J. Wang.
\newblock {Flow Vorticity in Peripheral High Energy Heavy Ion Collisions}.
\newblock {\em Phys. Rev. C}, 87(3):034906, 2013.

\bibitem{Csernai:2014ywa}
L.P. Csernai, D.J. Wang, M.~Bleicher, and H.~Stoecker.
\newblock {Vorticity in peripheral collisions at the Facility for Antiproton
  and Ion Research and at the JINR Nuclotron-based Ion Collider fAcility}.
\newblock {\em Phys. Rev. C}, 90(2):021904, 2014.

\bibitem{Becattini:2015ska}
F.~Becattini, G.~Inghirami, V.~Rolando, A.~Beraudo, L.~Del~Zanna, A.~De~Pace,
  M.~Nardi, G.~Pagliara, and V.~Chandra.
\newblock {A study of vorticity formation in high energy nuclear collisions}.
\newblock {\em Eur. Phys. J. C}, 75(9):406, 2015.
\newblock [Erratum: Eur.Phys.J.C 78, 354 (2018)].

\bibitem{Teryaev:2015gxa}
Oleg Teryaev and Rahim Usubov.
\newblock {Vorticity and hydrodynamic helicity in heavy-ion collisions in the
  hadron-string dynamics model}.
\newblock {\em Phys. Rev. C}, 92(1):014906, 2015.

\bibitem{Jiang:2016woz}
Yin Jiang, Zi-Wei Lin, and Jinfeng Liao.
\newblock {Rotating quark-gluon plasma in relativistic heavy ion collisions}.
\newblock {\em Phys. Rev. C}, 94(4):044910, 2016.
\newblock [Erratum: Phys.Rev.C 95, 049904 (2017)].

\bibitem{Deng:2016gyh}
Wei-Tian Deng and Xu-Guang Huang.
\newblock {Vorticity in Heavy-Ion Collisions}.
\newblock {\em Phys. Rev. C}, 93(6):064907, 2016.

\bibitem{Ivanov:2017dff}
Yu.~B. Ivanov and A.A. Soldatov.
\newblock {Vorticity in heavy-ion collisions at the JINR Nuclotron-based Ion
  Collider fAcility}.
\newblock {\em Phys. Rev. C}, 95(5):054915, 2017.

\bibitem{Shi:2017wpk}
Shuzhe Shi, Kangle Li, and Jinfeng Liao.
\newblock {Searching for the Subatomic Swirls in the CuCu and CuAu Collisions}.
\newblock {\em Phys. Lett. B}, 788:409--413, 2019.

\bibitem{Becattini:2007nd}
F.~Becattini and F.~Piccinini.
\newblock {The Ideal relativistic spinning gas: Polarization and spectra}.
\newblock {\em Annals Phys.}, 323:2452--2473, 2008.

\bibitem{Becattini:2013fla}
F.~Becattini, V.~Chandra, L.~Del~Zanna, and E.~Grossi.
\newblock {Relativistic distribution function for particles with spin at local
  thermodynamical equilibrium}.
\newblock {\em Annals Phys.}, 338:32--49, 2013.

\bibitem{Zubarev_tmp1979_zps}
D.~N. Zubarev, A.~V. Prozorkevich, and S.~A. Smolyanskii.
\newblock Derivation of nonlinear generalized equations of quantum relativistic
  hydrodynamics.
\newblock {\em Theoretical and Mathematical Physics}, 40(3):821--831, Sep 1979.

\bibitem{Weert_ap1982}
Ch.G van Weert.
\newblock Maximum entropy principle and relativistic hydrodynamics.
\newblock {\em Annals of Physics}, 140(1):133 -- 162, 1982.

\bibitem{Becattini:2014yxa}
F.~Becattini, L.~Bucciantini, E.~Grossi, and L.~Tinti.
\newblock {Local thermodynamical equilibrium and the beta frame for a quantum
  relativistic fluid}.
\newblock {\em Eur. Phys. J. C}, 75(5):191, 2015.

\bibitem{Gao:2007bc}
Jian-Hua Gao, Shou-Wan Chen, Wei-tian Deng, Zuo-Tang Liang, Qun Wang, and
  Xin-Nian Wang.
\newblock {Global quark polarization in non-central A+A collisions}.
\newblock {\em Phys. Rev. C}, 77:044902, 2008.

\bibitem{Chen:2008wh}
Shou-wan Chen, Jian Deng, Jian-hua Gao, and Qun Wang.
\newblock {A General derivation of differential cross-section in quark-quark
  scatterings at fixed impact parameter}.
\newblock {\em Front. Phys. China}, 4:509--516, 2009.

\bibitem{Huang:2011ru}
Xu-Guang Huang, Pasi Huovinen, and Xin-Nian Wang.
\newblock {Quark Polarization in a Viscous Quark-Gluon Plasma}.
\newblock {\em Phys. Rev. C}, 84:054910, 2011.

\bibitem{Becattini:2009wh}
F.~Becattini and L.~Tinti.
\newblock {The Ideal relativistic rotating gas as a perfect fluid with spin}.
\newblock {\em Annals Phys.}, 325:1566--1594, 2010.

\bibitem{Becattini:2012tc}
F.~Becattini.
\newblock {Covariant statistical mechanics and the stress-energy tensor}.
\newblock {\em Phys. Rev. Lett.}, 108:244502, 2012.

\bibitem{Becattini:2015nva}
F.~Becattini and E.~Grossi.
\newblock {Quantum corrections to the stress-energy tensor in thermodynamic
  equilibrium with acceleration}.
\newblock {\em Phys. Rev. D}, 92:045037, 2015.

\bibitem{Hayata:2015lga}
Tomoya Hayata, Yoshimasa Hidaka, Toshifumi Noumi, and Masaru Hongo.
\newblock {Relativistic hydrodynamics from quantum field theory on the basis of
  the generalized Gibbs ensemble method}.
\newblock {\em Phys. Rev. D}, 92(6):065008, 2015.

\bibitem{Florkowski:2017ruc}
Wojciech Florkowski, Bengt Friman, Amaresh Jaiswal, and Enrico Speranza.
\newblock {Relativistic fluid dynamics with spin}.
\newblock {\em Phys. Rev. C}, 97(4):041901, 2018.

\bibitem{Florkowski:2017dyn}
Wojciech Florkowski, Bengt Friman, Amaresh Jaiswal, Radoslaw Ryblewski, and
  Enrico Speranza.
\newblock {Spin-dependent distribution functions for relativistic hydrodynamics
  of spin-1/2 particles}.
\newblock {\em Phys. Rev. D}, 97(11):116017, 2018.

\bibitem{Montenegro:2017lvf}
David Montenegro, Leonardo Tinti, and Giorgio Torrieri.
\newblock {Sound waves and vortices in a polarized relativistic fluid}.
\newblock {\em Phys. Rev. D}, 96(7):076016, 2017.

\bibitem{Montenegro:2017rbu}
David Montenegro, Leonardo Tinti, and Giorgio Torrieri.
\newblock {Ideal relativistic fluid limit for a medium with polarization}.
\newblock {\em Phys. Rev. D}, 96(5):056012, 2017.
\newblock [Addendum: Phys.Rev.D 96, 079901 (2017)].

\bibitem{Hattori:2019lfp}
Koichi Hattori, Masaru Hongo, Xu-Guang Huang, Mamoru Matsuo, and Hidetoshi
  Taya.
\newblock {Fate of spin polarization in a relativistic fluid: An
  entropy-current analysis}.
\newblock {\em Phys. Lett. B}, 795:100--106, 2019.

\bibitem{Gallegos:2021bzp}
A.~D. Gallegos, U.~G\"ursoy, and A.~Yarom.
\newblock {Hydrodynamics of spin currents}.
\newblock 1 2021.

\bibitem{Li:2020eon}
Shiyong Li, Mikhail~A. Stephanov, and Ho-Ung Yee.
\newblock {Non-dissipative second-order transport, spin, and pseudo-gauge
  transformations in hydrodynamics}.
\newblock 11 2020.

\bibitem{Bhadury:2020puc}
Samapan Bhadury, Wojciech Florkowski, Amaresh Jaiswal, Avdhesh Kumar, and
  Radoslaw Ryblewski.
\newblock {Relativistic dissipative spin dynamics in the relaxation time
  approximation}.
\newblock {\em Phys. Lett. B}, 814:136096, 2021.

\bibitem{Fukushima:2020ucl}
Kenji Fukushima and Shi Pu.
\newblock {Spin Hydrodynamics and Symmetric Energy-Momentum Tensors -- A
  current induced by the spin vorticity --}.
\newblock 10 2020.

\bibitem{Florkowski:2018fap}
Wojciech Florkowski, Radoslaw Ryblewski, and Avdhesh Kumar.
\newblock {Relativistic hydrodynamics for spin-polarized fluids}.
\newblock {\em Prog. Part. Nucl. Phys.}, 108:103709, 2019.

\bibitem{Speranza:2020ilk}
Enrico Speranza and Nora Weickgenannt.
\newblock {Spin tensor and pseudo-gauges: from nuclear collisions to
  gravitational physics}.
\newblock 6 2020.

\bibitem{Gao:2012ix}
Jian-Hua Gao, Zuo-Tang Liang, Shi Pu, Qun Wang, and Xin-Nian Wang.
\newblock {Chiral Anomaly and Local Polarization Effect from Quantum Kinetic
  Approach}.
\newblock {\em Phys. Rev. Lett.}, 109:232301, 2012.

\bibitem{Chen:2012ca}
Jiunn-Wei Chen, Shi Pu, Qun Wang, and Xin-Nian Wang.
\newblock {Berry Curvature and Four-Dimensional Monopoles in the Relativistic
  Chiral Kinetic Equation}.
\newblock {\em Phys. Rev. Lett.}, 110(26):262301, 2013.

\bibitem{Hidaka:2016yjf}
Yoshimasa Hidaka, Shi Pu, and Di-Lun Yang.
\newblock {Relativistic Chiral Kinetic Theory from Quantum Field Theories}.
\newblock {\em Phys. Rev. D}, 95(9):091901, 2017.

\bibitem{Gao:2017gfq}
Jian-hua Gao, Shi Pu, and Qun Wang.
\newblock {Covariant chiral kinetic equation in the Wigner function approach}.
\newblock {\em Phys. Rev. D}, 96(1):016002, 2017.

\bibitem{Gao:2018wmr}
Jian-Hua Gao, Zuo-Tang Liang, Qun Wang, and Xin-Nian Wang.
\newblock {Disentangling covariant Wigner functions for chiral fermions}.
\newblock {\em Phys. Rev. D}, 98(3):036019, 2018.

\bibitem{Huang:2018wdl}
Anping Huang, Shuzhe Shi, Yin Jiang, Jinfeng Liao, and Pengfei Zhuang.
\newblock {Complete and Consistent Chiral Transport from Wigner Function
  Formalism}.
\newblock {\em Phys. Rev. D}, 98(3):036010, 2018.

\bibitem{Carignano:2018gqt}
Stefano Carignano, Cristina Manuel, and Juan~M. Torres-Rincon.
\newblock {Consistent relativistic chiral kinetic theory: A derivation from
  on-shell effective field theory}.
\newblock {\em Phys. Rev. D}, 98(7):076005, 2018.

\bibitem{Liu:2018xip}
Yu-Chen Liu, Lan-Lan Gao, Kazuya Mameda, and Xu-Guang Huang.
\newblock {Chiral kinetic theory in curved spacetime}.
\newblock {\em Phys. Rev. D}, 99(8):085014, 2019.

\bibitem{Gao:2019zhk}
Jian-Hua Gao, Zuo-Tang Liang, and Qun Wang.
\newblock {Dirac sea and chiral anomaly in the quantum kinetic theory}.
\newblock {\em Phys. Rev. D}, 101(9):096015, 2020.

\bibitem{Yang:2020mtz}
Shi-Zheng Yang, Jian-Hua Gao, Zuo-Tang Liang, and Qun Wang.
\newblock {Second-order Charge Currents and Stress Tensor in Chiral System}.
\newblock 3 2020.

\bibitem{Hou:2020mqp}
Defu Hou and Shu Lin.
\newblock {Polarization Rotation of Chiral Fermions in Vortical Fluid}.
\newblock 8 2020.

\bibitem{Fang:2016vpj}
Ren-hong Fang, Long-gang Pang, Qun Wang, and Xin-nian Wang.
\newblock {Polarization of massive fermions in a vortical fluid}.
\newblock {\em Phys. Rev. C}, 94(2):024904, 2016.

\bibitem{Weickgenannt:2019dks}
Nora Weickgenannt, Xin-Li Sheng, Enrico Speranza, Qun Wang, and Dirk~H.
  Rischke.
\newblock {Kinetic theory for massive spin-1/2 particles from the
  Wigner-function formalism}.
\newblock {\em Phys. Rev.}, D100(5):056018, 2019.

\bibitem{Gao:2019znl}
Jian-Hua Gao and Zuo-Tang Liang.
\newblock {Relativistic Quantum Kinetic Theory for Massive Fermions and Spin
  Effects}.
\newblock {\em Phys. Rev. D}, 100(5):056021, 2019.

\bibitem{Hattori:2019ahi}
Koichi Hattori, Yoshimasa Hidaka, and Di-Lun Yang.
\newblock {Axial Kinetic Theory and Spin Transport for Fermions with Arbitrary
  Mass}.
\newblock {\em Phys. Rev. D}, 100(9):096011, 2019.

\bibitem{Wang:2019moi}
Ziyue Wang, Xingyu Guo, Shuzhe Shi, and Pengfei Zhuang.
\newblock {Mass Correction to Chiral Kinetic Equations}.
\newblock {\em Phys. Rev. D}, 100(1):014015, 2019.

\bibitem{Liu:2020flb}
Yu-Chen Liu, Kazuya Mameda, and Xu-Guang Huang.
\newblock {Covariant Spin Kinetic Theory I: Collisionless Limit}.
\newblock 2 2020.

\bibitem{Becattini:2018duy}
F.~Becattini, Wojciech Florkowski, and Enrico Speranza.
\newblock {Spin tensor and its role in non-equilibrium thermodynamics}.
\newblock {\em Phys. Lett. B}, 789:419--425, 2019.

\bibitem{Florkowski:2018ahw}
Wojciech Florkowski, Avdhesh Kumar, and Radoslaw Ryblewski.
\newblock {Thermodynamic versus kinetic approach to polarization-vorticity
  coupling}.
\newblock {\em Phys. Rev. C}, 98(4):044906, 2018.

\bibitem{Li:2019qkf}
Shiyong Li and Ho-Ung Yee.
\newblock {Quantum Kinetic Theory of Spin Polarization of Massive Quarks in
  Perturbative QCD: Leading Log}.
\newblock {\em Phys. Rev. D}, 100(5):056022, 2019.

\bibitem{Kapusta:2020npk}
Joseph~I. Kapusta, Ermal Rrapaj, and Serge Rudaz.
\newblock {Spin versus Helicity Equilibration Times and Lagrangian for Strange
  Quarks in Rotating Quark-Gluon Plasma}.
\newblock 4 2020.

\bibitem{Zhang:2019xya}
Jun-jie Zhang, Ren-hong Fang, Qun Wang, and Xin-Nian Wang.
\newblock {A microscopic description for polarization in particle scatterings}.
\newblock {\em Phys. Rev. C}, 100(6):064904, 2019.

\bibitem{Yang:2020hri}
Di-Lun Yang, Koichi Hattori, and Yoshimasa Hidaka.
\newblock {Effective quantum kinetic theory for spin transport of fermions with
  collsional effects}.
\newblock 2 2020.

\bibitem{Weickgenannt:2020aaf}
Nora Weickgenannt, Enrico Speranza, Xin-li Sheng, Qun Wang, and Dirk~H.
  Rischke.
\newblock {Generating spin polarization from vorticity through nonlocal
  collisions}.
\newblock 5 2020.

\bibitem{Weickgenannt:2021cuo}
Nora Weickgenannt, Enrico Speranza, Xin-li Sheng, Qun Wang, and Dirk~H.
  Rischke.
\newblock {Derivation of the nonlocal collision term in the relativistic
  Boltzmann equation for massive spin-1/2 particles from quantum field theory}.
\newblock 3 2021.

\bibitem{DeGroot:1980dk}
S.R. De~Groot.
\newblock {\em {Relativistic Kinetic Theory. Principles and Applications}}.
\newblock 1 1980.

\bibitem{Martin:1959jp}
Paul~C. Martin and Julian~S. Schwinger.
\newblock {Theory of many particle systems. 1.}
\newblock {\em Phys. Rev.}, 115:1342--1373, 1959.
\newblock [,427(1959)].

\bibitem{Keldysh:1964ud}
L.~V. Keldysh.
\newblock {Diagram technique for nonequilibrium processes}.
\newblock {\em Zh. Eksp. Teor. Fiz.}, 47:1515--1527, 1964.
\newblock [Sov. Phys. JETP20,1018(1965)].

\bibitem{Chou:1984es}
Kuang-chao Chou, Zhao-bin Su, Bai-lin Hao, and Lu~Yu.
\newblock {Equilibrium and Nonequilibrium Formalisms Made Unified}.
\newblock {\em Phys. Rept.}, 118:1--131, 1985.

\bibitem{Blaizot:2001nr}
Jean-Paul Blaizot and Edmond Iancu.
\newblock {The Quark gluon plasma: Collective dynamics and hard thermal loops}.
\newblock {\em Phys. Rept.}, 359:355--528, 2002.

\bibitem{Berges:2004yj}
Juergen Berges.
\newblock {Introduction to nonequilibrium quantum field theory}.
\newblock {\em AIP Conf. Proc.}, 739(1):3--62, 2004.

\bibitem{Crossley:2015evo}
Michael Crossley, Paolo Glorioso, and Hong Liu.
\newblock {Effective field theory of dissipative fluids}.
\newblock {\em JHEP}, 09:095, 2017.

\bibitem{Nambu:1961tp}
Yoichiro Nambu and G.~Jona-Lasinio.
\newblock {Dynamical Model of Elementary Particles Based on an Analogy with
  Superconductivity. 1.}
\newblock {\em Phys. Rev.}, 122:345--358, 1961.

\bibitem{Nambu:1961fr}
Yoichiro Nambu and G.~Jona-Lasinio.
\newblock {Dynamical Model of Elementary Particles Based on an Analogy with
  Superconductivity. 2.}
\newblock {\em Phys. Rev.}, 124:246--254, 1961.

\bibitem{Kadanoff:1962}
L.~P. Kadanoff and G.~Baym.
\newblock {\em Quantum Statistical Mechanics}.
\newblock Benjamin, New York, 1962.

\bibitem{Itzykson:1980rh}
C.~Itzykson and J.B. Zuber.
\newblock {\em {Quantum Field Theory}}.
\newblock International Series In Pure and Applied Physics. McGraw-Hill, New
  York, 1980.

\bibitem{Mrowczynski:1992hq}
Stanislaw Mrowczynski and Ulrich~W. Heinz.
\newblock {Towards a relativistic transport theory of nuclear matter}.
\newblock {\em Annals Phys.}, 229:1--54, 1994.

\bibitem{Schonhofen:1994zf}
M.~Schonhofen, M.~Cubero, B.~L. Friman, W.~Norenberg, and G.~Wolf.
\newblock {Covariant kinetic equations and relaxation processes in relativistic
  heavy ion collisions}.
\newblock {\em Nucl. Phys.}, A572:112--140, 1994.

\bibitem{Vasak:1987um}
D.~Vasak, M.~Gyulassy, and H.~T. Elze.
\newblock {Quantum Transport Theory for Abelian Plasmas}.
\newblock {\em Annals Phys.}, 173:462--492, 1987.

\end{thebibliography}

\end{document}